\documentclass[%
 aip,
 amsmath,amssymb,
 reprint,%
]{revtex4-1}

\def\ispreprint{1}

\usepackage{graphicx}
\usepackage{dcolumn}
\usepackage{placeins}

\usepackage[utf8]{inputenc}
\usepackage[T1]{fontenc}
\usepackage{mathptmx}
\usepackage{etoolbox}
\usepackage{xcolor}
\usepackage{siunitx}
\usepackage{amsfonts}
\usepackage{amsmath}
\usepackage[textsize=tiny]{todonotes}
\usepackage[version=3]{mhchem}
\usepackage{mathtools}
\usepackage{float}
\usepackage{bbm}
\usepackage{bm}

\newcommand{\paragraphtitle}[1]{{\small\textsf{\textbf{#1}}} \normalfont}
\newcommand{\change}[1]{#1}
\newcommand{\changeR}[1]{{ #1}}

\usepackage{changes}
\makeatother

\begin{document}

\preprint{AIP/123-QED}

\title{Descriptors-free Collective Variables From Geometric Graph Neural Networks.}

\author{Jintu Zhang}
\thanks{Jintu Zhang and Luigi Bonati contributed equally to this work.}
\affiliation{Innovation Institute for Artificial Intelligence in Medicine of Zhejiang University, College of Pharmaceutical Sciences, Zhejiang University, Hangzhou 310058 Zhejiang, China}
\affiliation{Atomistic Simulations, Italian Institute of Technology, Genova 16152, Italy}

\author{Luigi Bonati}
\thanks{Jintu Zhang and Luigi Bonati contributed equally to this work.}
\affiliation{Atomistic Simulations, Italian Institute of Technology, Genova 16152, Italy}

\author{Enrico Trizio}
\affiliation{Atomistic Simulations, Italian Institute of Technology, Genova 16152, Italy}

\author{Odin Zhang}
\affiliation{Innovation Institute for Artificial Intelligence in Medicine of Zhejiang University, College of Pharmaceutical Sciences, Zhejiang University, Hangzhou 310058 Zhejiang, China}

\author{Yu Kang}
\affiliation{Innovation Institute for Artificial Intelligence in Medicine of Zhejiang University, College of Pharmaceutical Sciences, Zhejiang University, Hangzhou 310058 Zhejiang, China}

\author{TingJun Hou$^*$}
\email{tingjunhou@zju.edu.cn}
\affiliation{Innovation Institute for Artificial Intelligence in Medicine of Zhejiang University, College of Pharmaceutical Sciences, Zhejiang University, Hangzhou 310058 Zhejiang, China}
\affiliation{State Key Lab of CAD\&CG, Zhejiang University, Hangzhou, Zhejiang 310058, China}

\author{Michele Parrinello$^*$}
\email{michele.parrinello@iit.it}
\affiliation{Atomistic Simulations, Italian Institute of Technology, Genova 16152, Italy}


\begin{abstract}
Enhanced sampling simulations make the computational study of rare events feasible. 
A large family of such methods crucially depends on the 
definition of some collective variables (CVs) that could provide a low-dimensional representation of the relevant physics of the process.
Recently, many methods have been proposed to semi-automatize the CV design by using machine learning tools to learn the variables directly from the simulation data. 
However, most methods are based on feed-forward neural networks and require as input some user-defined physical descriptors.
Here, we propose to bypass this step using a graph neural network to directly use the atomic coordinates as input for the CV model.
This way, we achieve a fully automatic approach to CV determination that provides variables invariant under the relevant symmetries, especially the permutational one.
Furthermore, we provide different analysis tools to favor the physical interpretation of the final CV.
We prove the robustness of our approach using different methods from the literature for the optimization of the CV, and we prove its efficacy on several systems, including a small peptide, an ion dissociation in explicit solvent, and a simple chemical reaction.
\end{abstract}

\maketitle
\if\ispreprint1

\begin{figure}[b!]\centering
\includegraphics{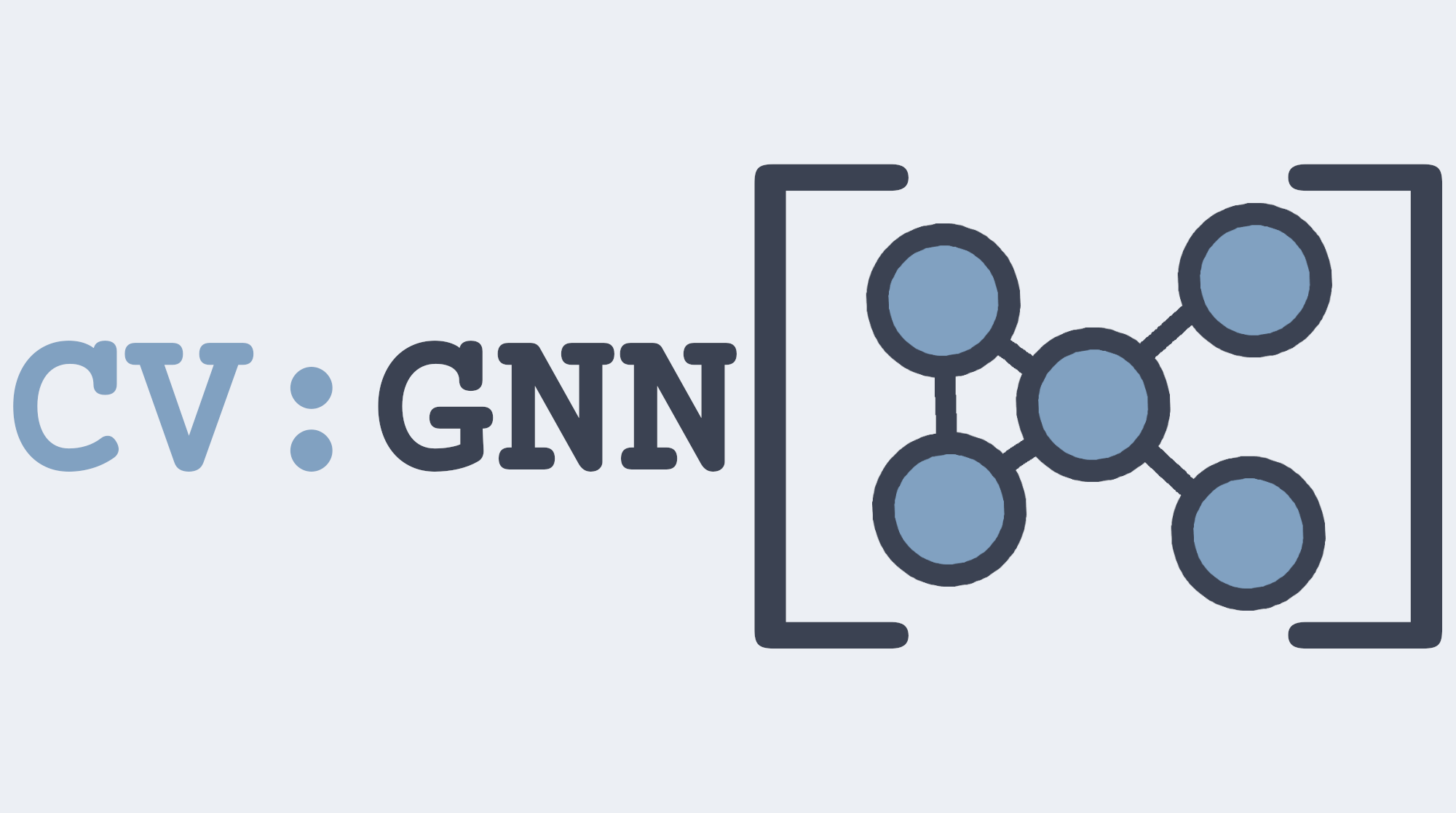}
\end{figure}
\else

\fi

\section{Introduction}

Molecular dynamics (MD) is a standard protocol for sampling molecular processes, which are most often dominated by transitions between different metastable conformation space states.
In the underlying free energy landscape, such states are found at local minima and are separated by high free energy barriers that hinder transitions.
Consequently, the barrier-crossing reactive events are rare compared to the timescales accessible to standard MD simulations.

To resolve the insufficient sampling caused by the presence of free energy barriers, a variety of enhanced sampling methods have been proposed~\cite{schulten2015,gao2019} to mitigate the effect of these barriers.
An important class of such methods, including at least Umbrella Sampling~\cite{torrie1977} and Metadynamics,~\cite{parrinello2002} makes use of the addition of an external biasing potential whose purpose is to accelerate sampling.
This makes possible computing both static and dynamic properties of the investigated process (e.g., free energies and kinetics)~\cite{parrinello20201,parrinello20231} within a moderate simulation time.

To keep the computational costs low and favor the interpretability of the results, the biasing potentials are typically applied to a small set of collective variables (CVs).
The CVs are functions of atomic positions and provide a low-dimensional representation of the transition process.
The performance of CV-based enhanced sampling methods heavily relies on the choice of CVs, and thus, the rational design of efficient CVs is one of the central problems in CV-based enhanced sampling.
In general, an effective CV has to be able to distinguish between the different conformation states and, possibly, to reflect the slow modes of the system.~\cite{noe2011,mcgibbon2017identification,chipot2023}

Traditionally, CVs have been chosen mostly based on chemical and physical intuition by selecting physical descriptors such as interatomic distances, (torsional) angles \change{or building more elaborate ones \cite{pietrucci2011,pietrucci2013}}.
However, it is well known that for complex systems, intuitively selected CVs may fail to satisfy the requirements mentioned above.~\cite{parrinello20221}
One strategy for resolving such a difficulty is to combine some of such primitive descriptors to obtain a CV that is more informative about the system.
Early implementations of this idea were based on simpler linear methods, such as Principal Component Analysis (PCA)~\cite{joliffe2016pca}, Time-lagged Independent Component Analysis (TICA)~\cite{noe2011} and Linear Discriminant Analysis (LDA)~\cite{parrinello2018}.
Unfortunately, in the case of complex systems, such approaches are limited by their linear nature, which, out of necessity, limits the flexibility and the number of input descriptors of the model.

Recently, to overcome this limitation, linear approaches have been replaced by the use of modern machine learning tools, such as artificial Neural Networks (NNs). In this way, larger sets of input descriptors can be non-linearly combined to obtain more flexible CVs.
Such Machine Learning CVs (MLCVs) are indeed much more expressive than those obtained with conventional linear methods and have already proved effective for various systems~\cite{trizio2024advancedsimulationsplumedopes}.
Conveniently, following the well-established patterns in the ML community, the optimization of NN-based CVs can be accomplished by imposing different training objectives.
For instance, supervised~\cite{parrinello20223,parrinello20203} and unsupervised~\cite{parrinello20232,andrew2018,stoltz2022,gervasio2024} learning methods have been used to build CVs primarily aimed at distinguishing different conformation states. 
Moreover, it has been shown that slow modes~\cite{parrinello20211,ferguson2019,tiwary2022} and committor probabilities~\cite{jung20231,parrinello2024,trizio2024onceprobabilitybasedenhancedsampling} could also be learned from appropriately designed loss functions. These objectives can be also effectively combined via multi-task learning~\cite{sun2022multitask, zhang2024combining}.

However, \change{a limitation of these methods comes from the need to choose a proper set of descriptors, which should include the relevant information about the transition, at least when considered collectively}.~\cite{trizio2024advancedsimulationsplumedopes}
Even with the possibility of using a larger number of inputs, this crucial choice mostly relies on prior knowledge about the system\changeR{, especially when physical descriptors are used as inputs}.
\changeR{A possible alternative to alleviate this issue is to draw from the several descriptors that have been suggested within the context of machine learning potentials. 
Such descriptors (or atomic fingerprints) are indeed designed to be general and include, for example, atom centered symmetry functions (ACSF)~\cite{parrinello20071}, permutationally invariant polynomials (PIP)~\cite{bowman2010}, smooth overlap of atomic position (SOAP)~\cite{bartok2013} and many others.
However, they require, for each system studied, setting up a number of parameters and also identifying a number of them out of a large pool of candidates \cite{imbalzano2018}. Above all, such methods scale poorly with the number of chemical species making it prohibitive to study,
for instance, biological systems.
Other attempts closer in spirit to the universal descriptors rely on graph theory to obtain appropriate descriptors, such as social permutation invariant coordinates (SPRINT)~\cite{pietrucci2011}, and/or pairwise interactions, like, for instance, permutation invariant vectors (PIV).~\cite{pietrucci2018}
However, they only encode scalar distances or contacts, and as such, they might provide an incomplete representation of the system. 

}

\changeR{Ideally, we would like the representation of the atomic system to be complete and at the same time "descriptors-free". In particular, by this, we mean not having to construct or select a subset of features for each system or process considered. To this extent, a possibility 
is to base the MLCVs model on geometric Graph Neural Networks (GNNs)~\cite{pietro2023} to directly take atomic coordinates as input.}
GNNs represent atomistic systems via geometric graphs, where atoms are represented by nodes, and the edges between nodes reflect their connection relationships.
Nodes and/or edges are also assigned attributes that can be used to encode spatial information on the system.
Another advantage of such architectures is that, by design, the scalar and vector features learned by the geometric GNNs can be made invariant and equivariant, respectively, \textit{w.r.t.} the overall translations and rotations of the graph, which allow the network to preserve the symmetries of atomistic systems.
Thus, in principle, well-designed geometric GNNs can learn most structural features of arbitrary atomistic systems without manually defining descriptors in advance.
Consequently, geometric GNNs are promising candidates to obtain expressive MLCVs without the need to build and select descriptors.

\change{Realizing the advantages of geometric GNNs over simpler feedforward networks, several works have been reported that exploit them to build MLCVs.
For instance, \citeauthor{salvalaglio2024} used a GNN architecture together with data augmentation to speed up the computation of local bond order parameters for sampling nucleation.~\cite{salvalaglio2024} 
Similarly, \citeauthor{tiwary20242} applied a roto-translation invariant geometric GNN to model phase transitions of crystals~\cite{tiwary20242} with the caveat of selecting specific edge features for different systems.}
\changeR{Following a different 
strategy, multitask approaches combining the learning of CVs with that of an interatomic potential have been proposed by \citeauthor{sipka2023}\cite{sipka2023}, using variational autoencoders, and \citeauthor{speybroeck2024}\cite{speybroeck2024}, via a classification criterion.}
\changeR{In addition, GNNs have also been applied to construct Markov state models~\cite{mohdi2022} and characterize molecular conformations for different purposes~\cite{noe2021}.
}

\change{
This very same approach is followed in this paper, but is made more general.
Indeed, we focus on using GNNs as universal featurizers in a flexible framework for the construction of CVs that can be applied to different physical and chemical systems and optimized on different criteria, from classification to slow mode extraction.}

Among the architectures presented in the literature, here, we adopt the Geometric Vector Perceptron (GVP).~\cite{dror20211,dror20212}
Our choice is motivated by the fact that the GVP architecture provides a good balance between computational complexity and accuracy, as proved by its use in protein design 
\changeR{and being equivariant to the \textit{E(3)} symmetry group, i.e., \textit{w.r.t.} overall translations, rotations, and reflections of the input geometry.}
\change{Nonetheless, thanks to the code interfaces we developed, different GNN choices could also be easily explored}\changeR{, for example, those equivariant under \textit{SE(3)} which can account for overall system chirality.~\cite{morehead2024}}

To show the generality and flexibility of our approach, we tested our GVP-GNN-based MLCVs using different learning objectives and different types of atomistic processes.
For the optimization of the CV models, we adopted the Deep Targeted Discriminant Analysis (DeepTDA)~\cite{parrinello20223} and Deep Time-Lagged Independent Component Analysis (DeepTICA)~\cite{parrinello20211} objective functions.
The first tested process is the conformation transition of alanine dipeptide in a vacuum, which is widely used in benchmarking enhanced sampling algorithms.
Then, we studied the dissociation of sodium chloride (\ce{NaCl}) in bulk water to examine if our approach could learn meaningful CVs from noisy training data.
Finally, we investigated the methyl migration of a 2-fluoro-2,3-dimethyl-butane (FDMB) cation and compared the performance of GNN-based MLCVs with that of feedforward network-based MLCVs, proving the importance of permutation invariance in building MLCVs.
We found that for each system, the CVs thus obtained are robust in biased free energy calculations.
Moreover, we also showed how to interpret the CV model to obtain physical insights into the conformation transitions.
Our results clearly demonstrate that GNN-based MLCVs can precisely capture the key features of various atomistic processes.
What's more, due to the universality of geometric GNNs, our methodology could be easily generalized to other training objectives, like, learning the committor probabilities.~\cite{parrinello2024, trizio2024onceprobabilitybasedenhancedsampling}

\section{Methods}

\if\ispreprint1
    \begin{figure*}[th!]\centering
    \includegraphics[width=\linewidth]{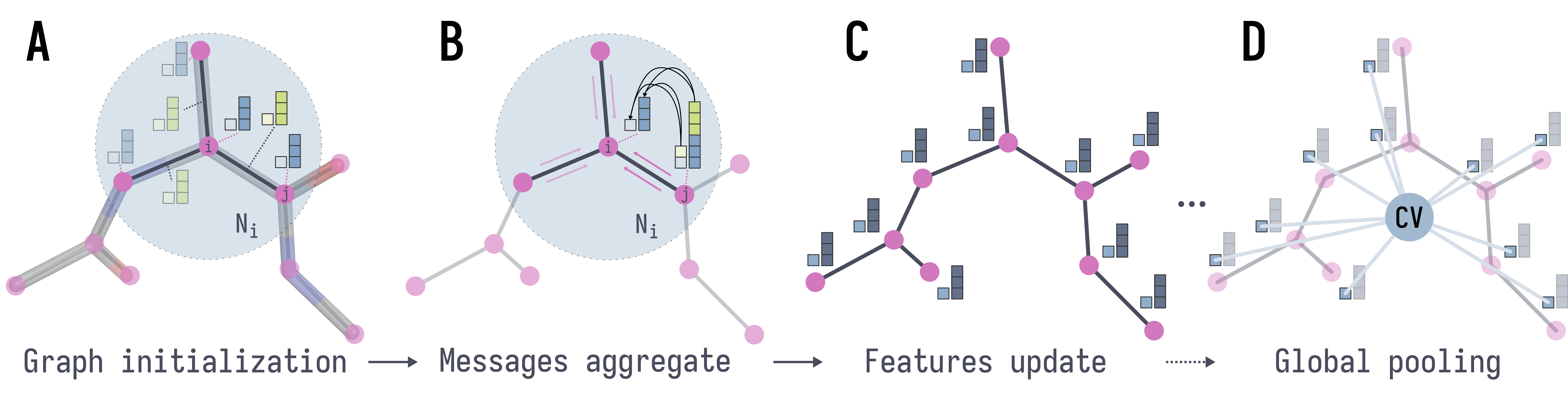}
    \caption{\change{Schematic representation of the framework for GVP-GNN-based collective variable (CV) using alanine dipeptide as an example.
    (A) The molecular structure is represented with a graph, whose nodes (pink dots) represent the atoms and the edges (grey lines) the spatial relationship between them. Both objects are associated with scalar and vector features, and the edges are determined based on a radial cutoff that determines the neighborhood $N_i$ of atom $i$ (shaded blue region).
    (B) Through the hidden layers of the network, the information is propagated following the message-passing scheme. For a node $i$, messages from the neighboring nodes $j$ are obtained by concatenating their scalar and vector features and those of the edge connecting them. The concatenated object is the processed with non-linear GVP layers to mix the scalar and vector channels. For clarity, we only depict this for a single $i$-$j$ pair.
    (C) The node features of the whole graph are updated based on the information gathered from the neighbors' messages.
    (D) The CV, which is a scalar and invariant quantity, is obtained by performing a global pooling of the scalar node features of the output layer of the network.  
    }}
    \label{fig:scheme}
    \end{figure*}
\else
    \begin{figure}\centering
    \includegraphics[width=\linewidth]{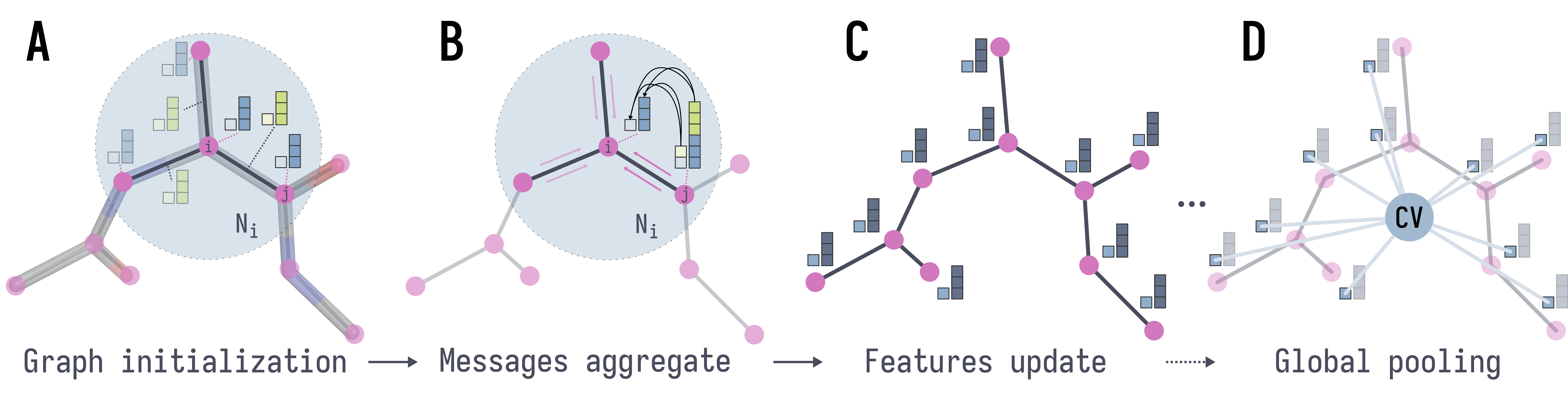}
    \caption{\change{Schematic representation of the framework for GVP-GNN collective variable (CV) using alanine dipeptide as an example.
    (A) The molecular structure is represented with a graph whose, nodes (pink dots) represent the atoms and the edges (grey lines) the spatial relationship between them. Both objects are associated with scalar and vector features, and the edges are determined based on a radial cutoff that determines the neighborhood $N_i$ of atom $i$ (shaded blue region).
    (B) Through the hidden layers of the network, the information is propagated following the message-passing scheme. For a node $i$, messages from the neighboring nodes $j$ are obtained by mixing the scalar and vector channels of the two nodes and the edge connecting them and then aggregated. For clarity, we only depict this for a single $ij$ pair.
    (C) The node features of the whole graph are updated based on the information gathered from the neighbors' messages.
    (D) The CV, which is a scalar and invariant quantity, is obtained by performing a global pooling of the scalar node features of the output layer of the network.  
    }}
    \label{fig:scheme}
    \end{figure}
\fi

\subsection{Geometric Graph Neural Networks.}

As anticipated in the introduction, geometric GNNs provide a natural platform for modeling atomistic systems.~\cite{duval2023hitchhiker}
\change{Here, we first summarize the basic elements of geometric graphs and GNNs and then provide an overview of the specific architecture we used in this work, still referring the reader to the original works for the full details.~\cite{dror20211, dror20212}}

\change{\paragraphtitle{Background.} In the geometric GNN framework}, the system is represented by geometric graphs, which are attributed graphs that contain a set of $n$ nodes and the edges connecting them.
\change{Formally, a geometric graph is expressed as $\mathcal{G} = \left(\bm{A}, \bm{S}, \vec{\bm{V}}, \vec{\bm{X}}\right)$, where $\bm{A}$ and $\bm{S}$ stand for the adjacency matrix and scalar node features and $\vec{\bm{V}}$ and $\vec{\bm{X}}$ represent the vector node features and node positions.
Additional information such as distances between nodes can be stored in scalar $\bm{S}_e$ and vector $\vec{\bm{V}_e}$ edge features (see Figure~\ref{fig:scheme}A). }

In geometric GNNs, the information is typically propagated through the network layers following the message-passing (MP) scheme,~\cite{dahl2017} in which node features are updated using the information coming from \change{the node's} neighbors.
In the context of atomistic systems, this approach allows the network to capture the interactions based on the graph topology and the atomic coordinates.
In practice, the MP algorithm consists of two operations: the message aggregate and the node feature update.
In the message aggregate step, so-called ``messages'' are calculated between connected nodes using a learnable message function (\textsc{Agg}) according to their node features and relative spatial positions \change{(see Figure~\ref{fig:scheme}B)}.
Then, the update step is performed, in which a learnable update function  (\textsc{Upd}) updates the features of each node using the messages received from its neighbors \change{(see Figure~\ref{fig:scheme}C)}.

According to the type of message functions, geometric GNN models can be classified as invariant or equivariant based on their behavior \textit{w.r.t.} symmetry operations.
In invariant GNNs, for example, SchNet,~\cite{schutt2018} messages between nodes only contain scalar quantities and are thus invariant \textit{w.r.t.} rotations and translations of the entire graph.
On the other hand, in equivariant GNNs, e.g., GVP-GNN~\cite{dror20211,dror20212} and MACE,~\cite{grabor2022} messages between nodes contain additional vector channels that are equivariant \textit{w.r.t.} to rotations and translations.
As equivariant GNNs contain higher-order geometric information, they are typically more expressive and flexible than their invariant counterparts while also being more data-efficient.~\cite{pietro2023} \change{At the same time, they can be used to learn also invariant quantities (see Sec.~\ref{sec:GNN-MLCVS}).}
Thus, here we focus on equivariant GNNs, which\change{, in general, can be described based on the following two operations~\cite{pietro2023}:} 
\begin{equation}
\begin{split}
& \bm{s}_{i}^m, \vec{\bm{v}}_{i}^m \coloneq \textsc{Agg} \left( \left\{\!\left\{ \left(\bm{s}_i^o, \vec{\bm{v}}_i^o\right), \left(\bm{s}_j^o, \vec{\bm{v}}_j^o\right), \vec{\bm{x}}_{ij} \big| j \in \mathcal{N}_i \right\}\!\right\} \right) \\
& \bm{s}_i^n, \vec{\bm{v}}_i^n \leftarrow \textsc{Upd} \left( \left(\bm{s}_i^o, \vec{\bm{v}}_i^o\right),\left(\bm{s}_i^m, \vec{\bm{v}}_i^m\right)\right). \label{eqn:1}
\end{split}
\end{equation}
In the above equations, $\left(\bm{s}_i^o, \vec{\bm{v}}_i^o\right)$ \change{stands for the old scalar and vector features}, $\left(\bm{s}_i^n, \vec{\bm{v}}_i^n\right)$ \change{for the new scalar and vector features} and $\left(\bm{s}_i^m, \vec{\bm{v}}_i^m\right)$ \change{for} the messages incoming into a given node $i$.
$\vec{\bm{x}}_{ij}$ represents the relative position vector between node $i$ and one of \change{the nodes $j$ from its set of neighbors $\mathcal{N}_i$.}

\change{\paragraphtitle{Geometric Vector Perceptron.}}
In this study, we chose the vector-gated GVP model \change{proposed in Ref.~\citenum{dror20212}} as GNN architecture, which provides a balanced trade-off between computational costs and expressivity.
\change{In GVP-GNN, the geometric information of the system is based on node distances, which are stored in the edge features.
In particular, edge scalar features $\bm{s}_e$ are the distances expanded by a set of radial bases,~\cite{dahl2017, gasteiger2022} while the vector features $\vec{\bm{v}}_e$ correspond to normalized orientation vectors.
On the other hand, scalar node features $\bm{s}^0$ are initialized as a one-hot encoding of the atomic types, and vector node features $\vec{\bm{v}}^0$ are simply initialized as zeros.}

\change{\paragraphtitle{GVP layer.} The scalar and vector channels are transformed equivariantly using the non-linear GVP layers.
Each GVP layer, takes as input a set of old scalar and vector features $\bm{s}^o$ and $\vec{\bm{v}}^o$, and transform them into new scalar node features $\bm{s}^n$ as:
    \begin{equation}
        \bm{s}^n \leftarrow \sigma \left(\bm{s}'\right) \quad\text{with}\quad  \bm{s}' \coloneq \bm{W}_m \left[{\|\bm{W}_h\vec{\bm{v}}^o\|_2 \atop \bm{s}^o}\right] + \bm{b}
        \label{eqn:2}
    \end{equation}
whereas new vector \change{node} features $\vec{\bm{v}}^n$ are computed as:
    \begin{equation}
        \vec{\bm{v}}^n \leftarrow \sigma_g \left(\bm{W}_g\left(\sigma^+ \left(\bm{s}'\right)\right) + \bm{b}_g \right) \odot \bm{W}_\mu\bm{W}_h\vec{\bm{v}}^o.
        \label{eqn:3}
    \end{equation}
}
\change{The GVP utilizes distinct linear learnable transformations ($\bm{W}_m$, $\bm{W}_\mu$, $\bm{W}_h$ and $\bm{W}_g$), learnable biases ($\bm{b}$ and $\bm{b}_g$) as well as non-linear activations ($\sigma$, $\sigma_g$ and $\sigma^+$) for updating the network's scalar and vector channels.
To ensure exchange of information between different types of channels, the $L_2$ norms of the partially updated vector features ($\|\bm{W}_h\vec{\bm{v}}^o\|_2$) are concatenated to the old scalar features ($\bm{s}^o)$.
This concatenated object is transformed into an intermediate scalar quantity $\bm{s}'$, which is used to update both scalar and vector features. 
Then, the new scalar features $\bm{s}^n$ are obtained simply by applying a non-linear activation $\sigma$ to $\bm{s}'$.
On the other hand, the new vector features $\vec{\bm{v}}^n$ are computed as the row-wise multiplication ($\odot$) of two terms. One is the vector gate ($\sigma_g \left(\bm{W}_g\left(\sigma^+ \left(\bm{s}'\right)\right) + \bm{b}_g \right)$) applied to $\bm{s}'$ and the other is a linear transformation of the old vector features ($\bm{W}_\mu\bm{W}_h\vec{\bm{v}}^o$).}

\paragraphtitle{Message passing.} 
In the GVP-GNN architecture, for a center node $i$, the messages $(\bm{s}^m_{(j\rightarrow i)},\vec{\bm{v}}^m_{(j\rightarrow i)})$
from the neighbors $j \in \mathcal{N}_i$ are constructed in two steps.
First, the features of node $i$ and $j$ are concatenated with those of the edge connecting them $(\bm{s}_e^{j \rightarrow i}, \vec{\bm{v}}_e^{j \rightarrow i})$, see Figure~\ref{fig:scheme}B.
Then, the combined features are turned into messages by feeding them to one or several GVP layers $g_1$.
The overall message function can thus be written in a compact form as:
\begin{equation}
    \bm{s}^m_{j\rightarrow i}, \vec{\bm{v}}^m_{j\rightarrow i} \coloneq g_1\left(
    \text{Concat}\left(
    (\bm{s}^o_j, \vec{\bm{v}}^o_j) , 
    (\bm{s}_e^{j \rightarrow i}, \vec{\bm{v}}_e^{j \rightarrow i}),
    (\bm{s}^o_i, \vec{\bm{v}}^o_i)
    \right)
    \right)
\end{equation}

The old features ($\bm{s}_i^o$,$\vec{\bm{v}}_i^o$) of node $i$ are then updated by adding the average value of the $k$ incoming messages:
\begin{equation}
    (\bm{s}_i^n,\vec{\bm{v}}_i^n) \leftarrow 
    (\bm{s}_i^o,\vec{\bm{v}}_i^o) + 
    \frac{1}{k}
    \sum_{j\in\mathcal{N}_i} \left(\bm{s}^m_{j\rightarrow i}, \vec{\bm{v}}^m_{j\rightarrow i}\right)
    \label{eq:mp}
\end{equation}

Furthermore, between the propagation steps, the information is processed applying node-wise GVP layer(s) $g_2$ to update the features of each node $i$:
\begin{equation}
    (\bm{s}_i^{\tilde{n}},\vec{\bm{v}}_i^{\tilde{n}}) \leftarrow 
    (\bm{s}_i^n,\vec{\bm{v}}_i^n) + g_2\left((\bm{s}_i^n,\vec{\bm{v}}_i^n)\right)
    \label{eq:nw_gvp}
\end{equation}
Note that an overall normalization and a dropout operation are also applied in both Equations~\ref{eq:mp} and \ref{eq:nw_gvp}.

\change{\paragraphtitle{Global pooling.}}
In practice, instead of analyzing individual node features directly, one is more interested in finding a global label to represent the status of the whole input graph\change{, which could be, for example, energies in the case of machine-learning potentials or, in our case, CVs}.
It is thus necessary to combine the information from different nodes into graph-level quantities.
As such collective quantities are generally invariant to the permutation of equivalent nodes, such a combination is often done by applying so-called permutation-invariant pooling functions, which can be, for example, a simple summation \change{or average} over the output \change{scalar} node features of the network (see Figure~\ref{fig:scheme}D).

\if\ispreprint1
    \begin{figure}[th!]
    \includegraphics[width=0.5\textwidth]{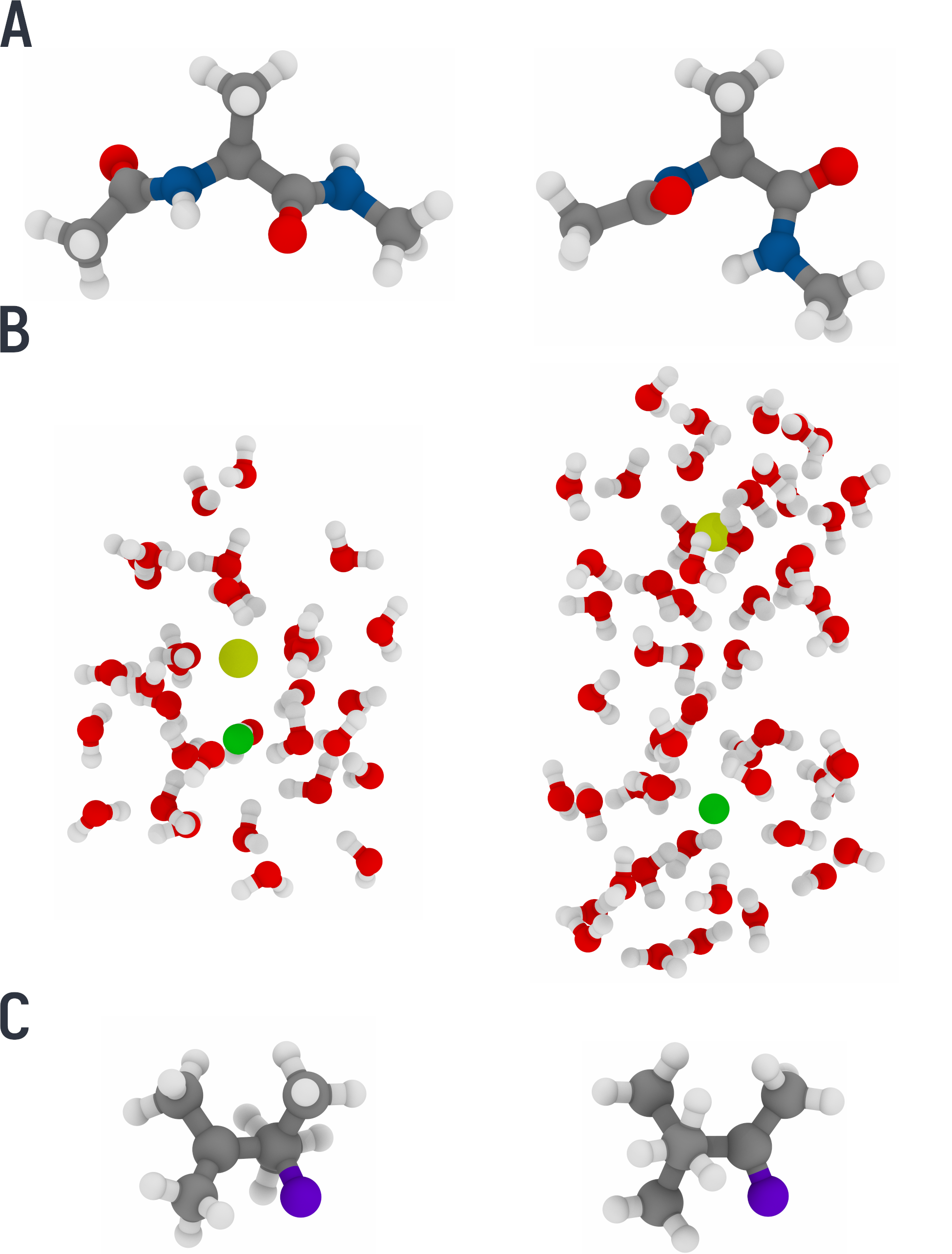}
    \caption{\changeR{Reactant (left panels) and product (right panels) state structures of the studied processes. 
    (A) Conformational equilibrium of alanine dipeptide in a vacuum. 
    (B) Dissociation of \ce{NaCl} in bulk water. For clarity, only water molecules within 0.6 \unit{\nm} from the ions are shown. 
    (C) Methyl migration in FDMB cation in vacuum.}}
    \label{fig:0}
    \end{figure}
\else
    \begin{figure}
    \includegraphics[width=0.5\textwidth]{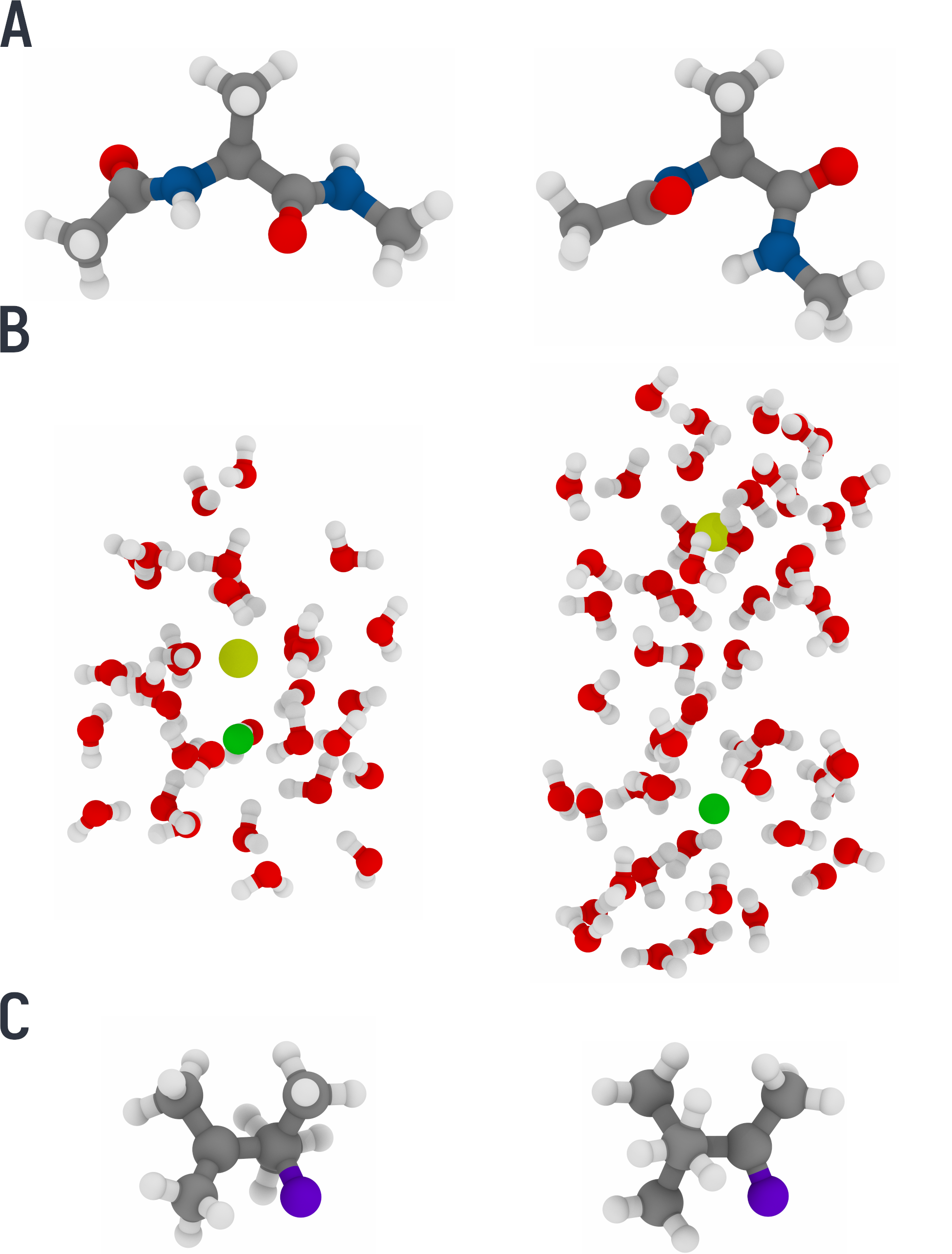}
    \caption{\changeR{Reactant (left panels) and product (right panels) state structures of the studied processes. 
    (A) Conformational equilibrium of alanine dipeptide in a vacuum. 
    (B) Dissociation of \ce{NaCl} in bulk water. For clarity, only water molecules within 0.6 \unit{\nm} from the ions are shown. 
    (C) Methyl migration in FDMB cation in vacuum.}}
    \label{fig:0}
    \end{figure}
\fi

\subsection{Machine Learning CVs Based On Geometric GNNs.}
\label{sec:GNN-MLCVS}
As mentioned in the Introduction, a good CV generally has to fulfill several requirements: to provide a low-dimensional representation of the process, to obey the symmetries of the system at hand, and to encode enough information about the process.
In the framework of MLCVs, the first requirement is addressed by expressing the CVs as learnable functions of the atomic positions that are optimized to map the original high-dimensional conformation space into a low-dimensional CV space.
Thus, as discussed in the previous section, (equivariant) geometric GNNs are good candidates for this type of task since their scalar outputs satisfy by design the roto-translation invariance, and permutation invariance can be ensured by performing a global pooling of them.

\change{
In practice, to build our GNN-based CVs we start by encoding the atomic system into an attributed graph in which the nodes represent the atoms and the edges connecting them are drawn based on a radial cutoff and represent their connection relationship (see Figure~\ref{fig:scheme}A).
This way, by employing neighbor lists, the cost for the graph construction scales linearly with the number of atoms.

As mentioned in the previous section, we initialize the node scalar features based on a one-hot encoding of the chemical species and the vector ones with zeros.
The edges' scalar features are the interatomic distances expanded on Bessel functions modulated by a smooth polynomial cutoff,~\cite{gasteiger2022} whereas the vector features correspond to the distance versors.

Then, we apply several GVP layers to process the information encoded in the graph and update its node features (see Figure~\ref{fig:scheme}B and C).
Finally, we obtain the CV value by performing an average pooling of the scalar node features of the output layer of the model (see Figure~\ref{fig:scheme}D).
It is worth noting that the number of resulting CVs is thus determined by the number of scalar node features in such a layer.

The model can then be optimized according to one of the learning objectives available in the already vast literature of MLCVs, such as dimensionality reduction, classification, or slow modes extraction~\cite{parrinello20232, trizio2024advancedsimulationsplumedopes}.
}
In practice, this corresponds to defining a loss function that mathematically formalizes such criterion and that, if minimized, guarantees that the CV has the desired properties.
In this study, to showcase the flexibility \change{and generality of a geometric GNN-based approach, we used for the different examples two different loss functions that were already employed in MLCVs based on feed-forward networks.
One is the classifier-like DeepTDA loss~\cite{parrinello20223, parrinello20233} and the other is the DeepTICA loss designed for slow modes extraction,~\cite{parrinello20211} which we summarize in the following.}

The idea behind the DeepTDA method~\cite{parrinello20223} is to obtain a meaningful CV space, expressed as the output of a \change{feed-forward} NN, where the metastable states are well distinguished.
By optimizing the TDA loss, one enforces that the data from the different metastable states in the training set, when projected in the CV space $z$, will be distributed according to a simple target distribution in which the states are well-defined, such as a mixture of Gaussians.
In practice, the TDA loss function $\mathcal{L}_{\mathrm{DeepTDA}}$ enforces the mean values $\mu$ and variances $\sigma$ of the distributions of each of the $N_\mathrm{C}$ classes (states) in the to match the respective target values $\mu^{tg}$ and $\sigma^{tg}$:
\begin{equation}
\mathcal{L}_{\mathrm{DeepTDA}} = \sum_{c=1}^{N_\mathrm{C}}\left( \alpha\left|\mu_c(z)-\mu_c^{\mathrm{tg}}\right|^2 + \beta\left|\sigma_c(z)-\sigma_c^{\mathrm{tg}}\right|^2\right),
\label{eqn:4}
\end{equation}
where the $\alpha$ and $\beta$ hyperparameters roughly scale the two terms to the same order of magnitude.
\change{In this framework, we easily modify the original recipe to our purposes by simply switching the feed-forward NN with our GNN model and thus expressing the CV space $z$ as its pooled scalar output.}

On the other hand, the DeepTICA~\cite{parrinello20211} loss aims to find CVs that approximates the slowest modes in the system.
In the original TICA method, a generalized eigenvalue problem is solved to approximate the eigenvalues and eigenfunctions of the transfer operator, which describes the underlying dynamics of the system.\cite{noe2011}
This amounts to finding a linear combination of the input features such that the outputs are maximally autocorrelated, as measured by the eigenvalues  
$\lambda_i$.
In the DeepTICA method, the basis functions for the application of TICA are expressed as the output of a feed-forward NN. 
The loss function then reads as:
\begin{equation}
\mathcal{L}_{\mathrm{DeepTICA}} = - \sum_{i=1}^{N_\mathrm{CV}} \tilde{\lambda}_i^2, \label{eqn:5}
\end{equation}
where $\tilde{\lambda}_i$ is the $i$-th largest non-trivial eigenvalue.
Thus, the objective of such a loss function is to transform the descriptors such that the $N_\mathrm{CV}$ eigenfunctions reach the slowest modes in the system.

Here, we modify the original recipe by replacing the feed-forward network with a geometric GNN\change{, expressing the basis functions as its pooled scalar outputs}.
Thus, the resulting MLCVs, which are the eigenfunctions corresponding to the $N_\mathrm{CV}$ optimized eigenvalues, read as:
\begin{equation}
z_i = \sum_{j = 1}^{N_s} \alpha_{ij} \bm{s}^{g}_j. \label{eqn:6}
\end{equation}
In the above equation, $\alpha_{ij}$ are the components of the $i$-th eigenvector of the time-lagged generalized eigenvalue problem, and $\bm{s}^{g}$ is the output of the permutation-invariant readout function with size $N_s$.


\change{

As can be seen, the presented workflow is very general and does not depend on the specific system at hand, and can be applied with different optimization criteria.
Indeed, our methodology provides a true descriptors-free approach that helps make the CV design process more automatic.
In addition, leveraging the expression power of geometric GNNs, crucial information about the simulated systems could be learned from the optimized features.}

\subsection{Interpreting The CVs.}

Since MLCVs are effective but not very informative compressed representations of the conformational space, it is worthwhile to interpret the models to gain insight into the system studied.
In this work, we explored two different techniques to inspect the GNN CVs. The first is a sensitivity analysis to identify the most relevant nodes of the graph, while the second approximates the GNN outputs with interpretable linear models.

\if\ispreprint1
\paragraphtitle{Node Level Sensitivity Analysis.}
\else
\subsubsection{Node Level Sensitivity Analysis.}
\fi
The first analysis measures node sensitivity, that is, the relative importance of a given node to GNN-based MLCV, which can inform us about the dominant components of the system.
The idea is simple: if a node is important to the CV, then the CV should be more sensitive to the change of position of that node, and such sensitivity can be monitored by looking at the CV's derivative \textit{w.r.t.} the node's Cartesian position.
Thus, giving a dataset with $n_g$ configurations (or graphs), the sensitivity of the $i$-th node reads as:
\begin{equation}
\mathbbm{s}_i = \frac{1}{n_g}\sum_{j=1}^{n_g}\left|\frac{\partial z\left(\mathcal{G}_j\right)}{\partial{\vec{\bm{X}}_{ji}}}\right|. \label{eqn:7}
\end{equation}
In the above equation, $\mathcal{G}_j$ is the $j$-th entry of the dataset, and $\vec{\bm{X}}_{ji}$ is the position of the $i$-th node in $\mathcal{G}_j$.

\if\ispreprint1
\paragraphtitle{Approximating The GNN With Sparse Linear Models.}
\else
\subsubsection{Approximating The GNN With Sparse Linear Models.}
\fi
Another strategy for understanding what MLCVs have learned from the data is to approximate them with simpler, more interpretable models. In particular, we can use sparse linear models:
\begin{equation}
f_{\bm{w}}(\bm{d}) = \sum_{i} {w}_i\ {d}_i\left(\vec{\bm{X}}\right), \label{eqn:8}
\end{equation}
where the coefficients $\bm{w}$ give an immediate measure of the significance of the input features $\bm{d}\left(\vec{\bm{X}}\right)$, which can be a list of physical descriptors evaluated on a given conformation $\vec{\bm{X}}$.
The sparsity condition, which means that only a few coefficients are different from zero, is essential to have a physically transparent model.
This can be obtained by Least Absolute Shrinkage and Selection Operator (LASSO) regression~\cite{tibshirani1996regression}, which introduces a penalty to the magnitude of the coefficients via an $L_1$ norm. 

While we recently used LASSO to characterize the metastable states by finding a minimal number of descriptors that are able to classify them,~\cite{parrinello20224} here we use it to approximate the output of the GNN-based MLCV $z$ with a linear model $f_{\bm{w}}$ using the following loss function:
\begin{equation}
    \mathcal{L}_{\mathrm{LASSO}}(\bm{w}) = \frac{1}{n_g} \sum_{i=1} ^{n_g} \left \Vert f_{\bm{w}}\left(\bm{d}\left(\vec{\bm{X}}_{i}\right)\right)-z\left(\mathcal{G}_i\right)\right \Vert^2+ \alpha\left \Vert \bm{w}\right \Vert_{1}, \label{eqn:9}
\end{equation}
where $\bm{d}\left(\vec{\bm{X}}_{i}\right)$ is the descriptor set evaluated on the $i$-th conformation of the dataset.
The regularization strength $\alpha$ measures the level of sparsity, \textit{i.e.}, how many coefficients are different from zero.
If we are able to find a linear expression that approximates (qualitatively) the GNN output with just a few terms, we obtain a transparent characterization of the model.
Certainly, the success of the result depends on the choice of the list of descriptors, which must be able to capture at least qualitatively the process under investigation.
Multiple descriptor lists (e.g., angles and distances) can be compared to find the representation that gives the highest score with a given number of features.
This way, we obtain a result similar to that of symbolic regression, but LASSO optimization is much simpler, and it allows for a very large number of descriptors to be used.
We note that any higher-order correlations can be easily incorporated by adding products between features to the list of descriptors as well.

\if\ispreprint1
    \begin{figure*}[th!]\centering
    \includegraphics[width=\linewidth]{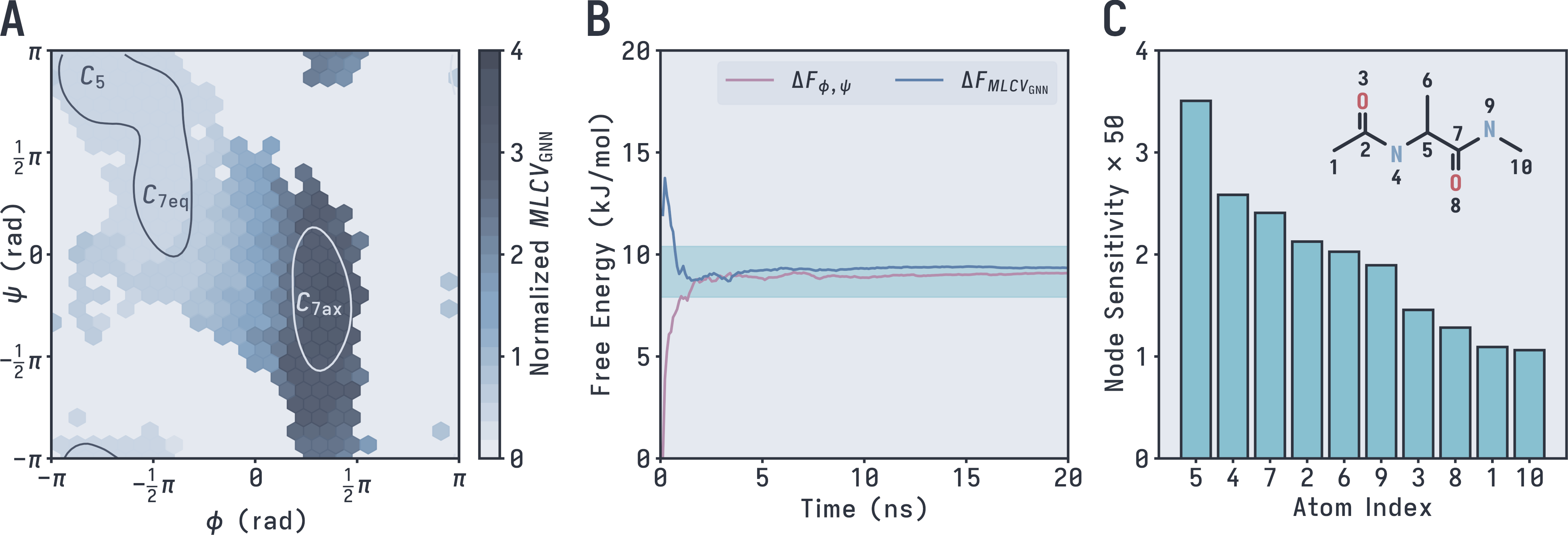}
    \caption{Results for the conformation transition of alanine dipeptide. (A) Values of the normalized MLCV projected on the $\phi$-$\psi$ space, with positions of different conformation states labeled. (B) Free energy difference between the initial and final conformation state during the OPES simulation. Results from both the MLCV and the torsional angles-based simulations are shown. The reference free energy $\pm 0.5 k_{\mathrm{B}}T$ range is represented by the colored region. (C) The node sensitivities calculated using Equation~\ref{eqn:7}. The dataset used to evaluate the sensitivities is the training set of the MLCV model.}
    \label{fig:1}
    \end{figure*}
\else
    \begin{figure}\centering
    \includegraphics[width=\linewidth]{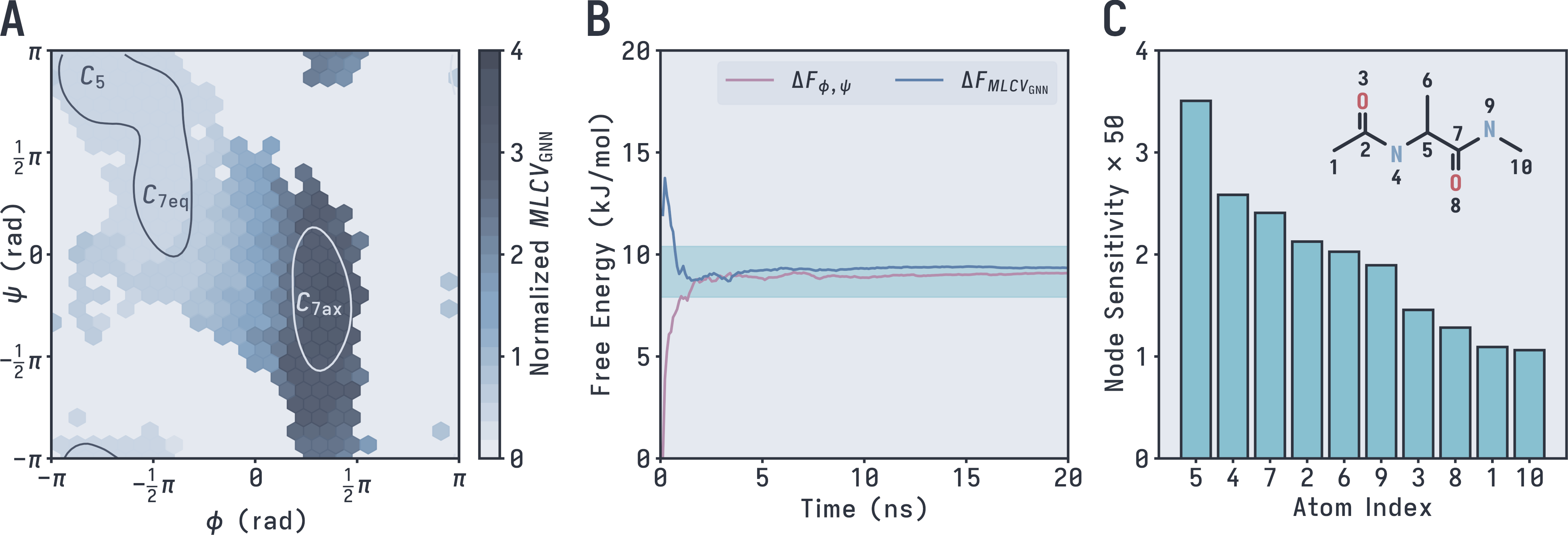}
    \caption{Results for the conformation transition of alanine dipeptide. (A) Values of the normalized MLCV projected on the $\phi$-$\psi$ space, with positions of different conformation states labeled. (B) Free energy difference between the initial and final conformation state during the OPES simulation. Results from both the MLCV and the torsional angles-based simulations are shown. The reference free energy $\pm 0.5 k_{\mathrm{B}}T$ range is represented by the colored region. (C) Node sensitivities, calculated evaluating Equation~\ref{eqn:7} on the training set of the MLCV model.}
    \label{fig:1}
    \end{figure}
\fi

\subsection{Training And Deploying CVs For Enhanced Sampling.}

The optimization of GNNs for enhanced sampling is carried out using a modified version of the \verb|mlcolvar| library,~\cite{parrinello20232, trizio2024advancedsimulationsplumedopes} a \verb|Python| toolkit specifically designed for training ML-based CVs with \verb|PyTorch|.
This approach facilitates the integration of various loss functions to optimize the CVs and provides access to several tools for interpreting the results, e.g., via sparse linear models. 

Once trained, the GNN-based CV is compiled using the PyTorch \verb|TorchScript| language, enabling deployment within the \verb|PLUMED| plugin for free energy and enhanced sampling calculations.~\cite{plumed} A modified version of the interface from \verb|pytorch| module of the plugin is used, which constructs the graph and computes the derivative of the CVs needed for the calculation of the additional forces through automatic differentiation. \change{Note that this interface is general so that the GVP layers could also be replaced with other convolutional layers with minimal modifications.} The integration of GNN-based CVs into \verb|PLUMED| offers significant flexibility, allowing their use with any available CV-based enhanced sampling method and MD engine.
\change{For the simulations presented here, we adopted the On-the-fly Probability Enhanced Sampling (OPES) method.~\cite{parrinello20202, parrinello20201,trizio2024advancedsimulationsplumedopes}}

\section{Results And Discussions}

\subsection{Alanine Dipeptide.}

As a first test for our method, we studied the well-known conformation transition of alanine dipeptide in a vacuum \changeR{(Figure~\ref{fig:0}A)} at 300 \unit{\kelvin}, which is a widely used model for benchmarking enhanced sampling methods.
To this end, we trained a GVP-GNN-based MLCV by optimizing the DeepTICA loss, with a lag time of 0.4 \unit{\ps}\change{, chosen such that the resulting TICA eigenvalues are neither too small nor too close to one~\cite{parrinello20211} and, in particular, this is the minimum value that produced eigenvalues of interest greater than 0.85.}
The dataset used for training the model was collected from an unbiased simulation trajectory generated at 600 \unit{\kelvin} and only contains the heavy atoms in the molecule.
To show the behavior of the MLCV, in Figure~\ref{fig:1}A, we plotted its values on the $\phi$-$\psi$ plane, in which the conformation states of the system are defined.
Clearly, the $C_\mathrm{5}$ and $C_{7\mathrm{ax}}$ states correspond to the minima and maxima of the MLCV, while the $C_{7\mathrm{eq}}$ state has a CV value slightly larger than that of the $C_\mathrm{5}$ state.
Meanwhile, the gradient direction of the MLCV is nearly parallel to the dominant $\phi$ angle, especially in the transition region ($-0.25\pi < \phi < 0.25\pi$).
These behaviors are consistent with those reported in other studies, which used a DeepTICA CV based on interatomic distances.~\cite{parrinello20211}

To assess the performances of the MLCV in the enhanced sampling context, we compared the results of two sets of On-the-fly Probability Enhanced Sampling (OPES)~\cite{parrinello20202} simulations, one performed by biasing the MLCV, the other using the conventional $\phi$ and $\psi$ as a reference.
In the MLCV-based OPES simulations, the free energy difference between the initial ($\phi<0$ \unit{\radian}) and final ($0$ \unit{\radian} $<\phi<2.2$ \unit{\radian}) states converged within $\pm 0.5 k_{\mathrm{B}}T$ from the reference value in less than 2 \unit{\ns} (blue line in Figure~\ref{fig:1}B).
The convergence speed is comparable to the one achieved in the torsional angles-based OPES simulation (see Figure~\ref{fig:1}B), and the sampling error estimated from three independent MLCV-based simulations is about 0.29 \unit{\kJ/\mole} (see Table~\ref{tbl:1}), which is relatively small.
Moreover, the free energy profile estimates from the two groups of simulations are practically indistinguishable (see Figure~S1A).
These results clearly exhibit the robustness of the MLCV in biased simulations.

Having assessed the efficacy of our graph-based and descriptors-free CV model, we shall now focus on extracting some physical knowledge from it.
To check what the MLCV has learned, we perform the node sensitivity analysis on the MLCV using Equation~\ref{eqn:7}.
As a result, we reassuringly found that, as expected, the most important atoms for the model are the four atoms involved in the dominant $\phi$ torsional angle (atoms 5, 4, 7, and 2 in Figure~\ref{fig:1}C).
Another confirmation of this result is obtained by performing a LASSO analysis of the four backbone torsional angles $\phi$, $\psi$, $\theta$, and $\omega$.
According to such analysis, the MLCV could be approximated by the following expression: $$MLCV \approx 0.926\phi + 0.06\theta - 0.014\psi,$$ and as expected, $\phi$ is found to be far more important than the other angles.
Interestingly, it is followed by $\theta$ (see Figure~S1B), consistently with finer analyses reported in previous studies.~\cite{parrinello2024,chandler2000}
In addition, we also carried out the LASSO analysis using the 45 distances between heavy atoms.
In this case, the most relevant descriptor this analysis suggests is the distance between atoms 3 and 6 (see Figure~S1C), which again agrees with previous work.~\cite{parrinello2024}
The great agreements between our results and those previously reported prove the capacity of the MLCV to capture finer, delicate details of the simulated systems while also providing an interpretable framework.

\if\ispreprint1
  \begin{table}[b!]
  \caption{Free energy difference between reactant and product states of each system. The errors for MLCV-based simulations are calculated from three independent simulations, and the errors for the reference values are estimated using the last 20\% of the reference simulation trajectories.}
  \label{tbl:1}
  \begin{tabular}{cccc}
    \hline
    System & Simulation Time& $\Delta F_{\mathrm{MLCV}}$ (\unit{\kJ/\mole}) & $\Delta F_{\mathrm{ref}}$ (\unit{\kJ/\mole})  \\
    \hline      Ala2 & 20 \unit{\ns} & 9.31 $\pm$ 0.29 & 9.11 $\pm$ 0.03 \\
    \hline \ce{NaCl} & 15 \unit{\ns} & -2.55 $\pm$ 0.14 & -2.87 $\pm$ 0.02 \\
    \hline      FDMB &  2 \unit{\ns} & 12.47 $\pm$ 0.23 & 12.67 $\pm$ 0.05 \\
    \hline
  \end{tabular}
  \end{table}
\else
  \begin{table}
  \caption{Free energy difference between reactant and product states of each system. The errors for MLCV-based simulations are calculated from three independent simulations, and the errors for the reference values are estimated using the last 20\% of the reference simulation trajectories.}
  \label{tbl:1}
  \begin{tabular}{cccc}
    \hline
    System & Simulation Length & $\Delta F$ (\unit{\kJ/\mole}) & $\Delta F_{\mathrm{ref}}$ (\unit{\kJ/\mole})  \\
    \hline      Ala2 & 20 \unit{\ns} & 9.31 $\pm$ 0.29 & 9.11 $\pm$ 0.03 \\
    \hline \ce{NaCl} & 15 \unit{\ns} & -2.55 $\pm$ 0.14 & -2.87 $\pm$ 0.02 \\
    \hline      FDMB &  2 \unit{\ns} & 12.47 $\pm$ 0.23 & 12.67 $\pm$ 0.05 \\
    \hline
  \end{tabular}
  \end{table}
\fi

\if\ispreprint1
    \begin{figure*}[th!]\centering
    \includegraphics[width=\linewidth]{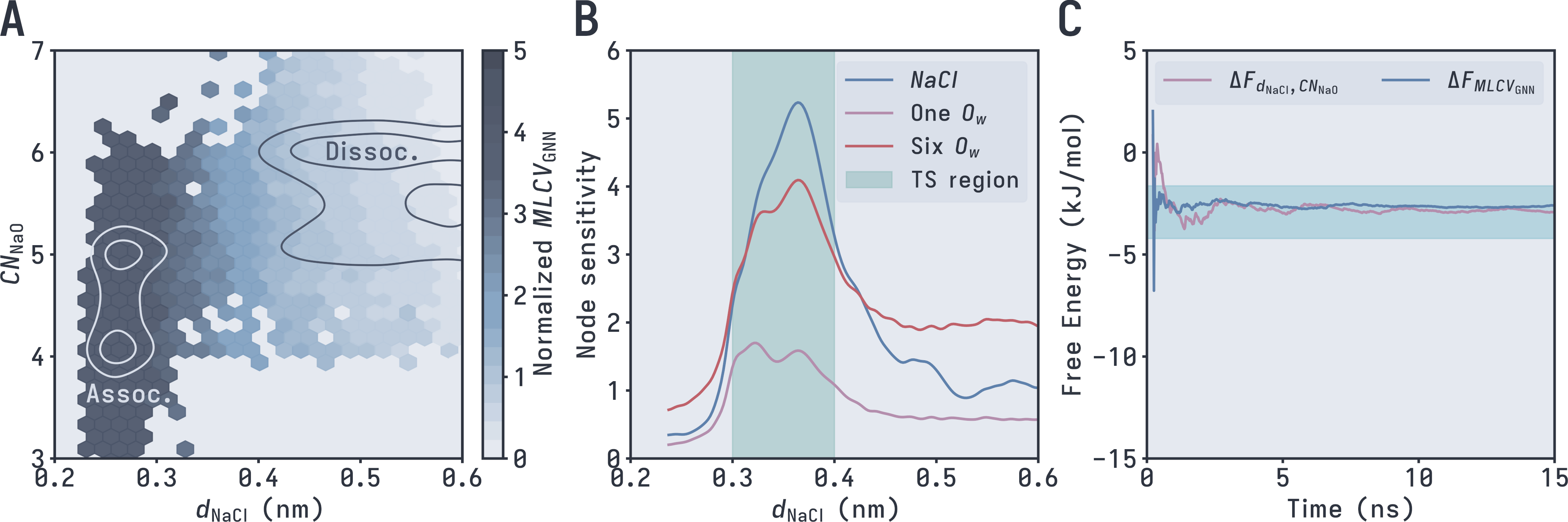}
    \caption{Results for \ce{NaCl} dissociation in water. (A) Normalized MLCV values projecting on the space spanned by $d_{\rm{NaCl}}$ and $CN_{\mathrm{NaO}}$. (B) Changing of max sensitivity of the different type of atoms \textit{w.r.t.} $d_{\rm{NaCl}}$. Here, the ``One $O_w$'' curve shows the sensitivity of the most sensitive oxygen atom, and the ``Six $O_w$'' curve shows the cumulative sensitivity of six most sensitive oxygen atoms. (C) The free energy difference between the two conformation states during the OPES simulation. Results from both the MLCV and classical CV-based simulations are shown. The reference free energy $\pm 0.5 k_{\mathrm{B}}T$ range is represented by the colored region.}
    \label{fig:2}
    \end{figure*}
\else
    \begin{figure}\centering
    \includegraphics[width=\linewidth]{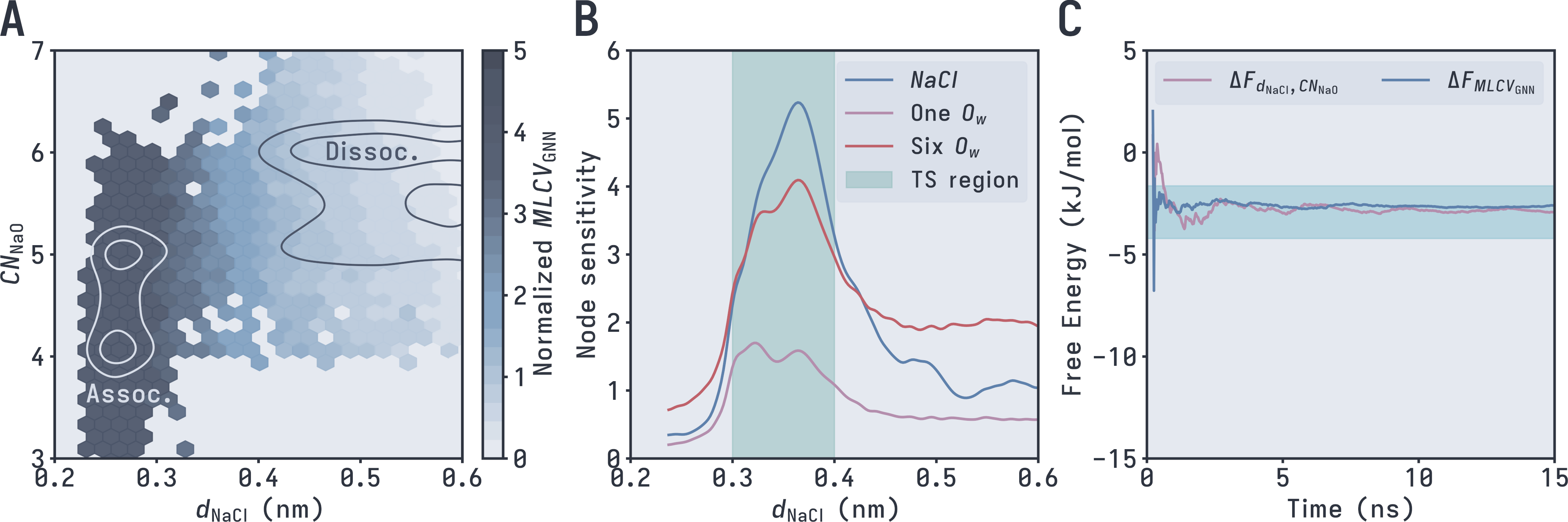}
    \caption{Results for \ce{NaCl} dissociation in water. (A) Normalized MLCV values projecting on the space spanned by $d_{\rm{NaCl}}$ and $CN_{\mathrm{NaO}}$. (B) Changing of max sensitivity of the different type of atoms \textit{w.r.t.} $d_{\rm{NaCl}}$. Here, the ``One $O_w$'' curve shows the sensitivity of the most sensitive oxygen atom, and the ``Six $O_w$'' curve shows the cumulative sensitivity of six most sensitive oxygen atoms. (C) The free energy difference between the two conformation states during the OPES simulation. Results from both the MLCV and classical CV-based simulations are shown. The reference free energy $\pm 0.5 k_{\mathrm{B}}T$ range is represented by the colored region.}
    \label{fig:2}
    \end{figure}
\fi

\subsection{\ce{NaCl} Dissociation In Water.}

As a second test for our method, we investigated the dissociation of a \ce{NaCl} ion pair in bulk water \changeR{(Figure~\ref{fig:0}B)} at 300 \unit{\kelvin} to examine whether our CV model could learn meaningful representations of solvated systems from noisy training data.
To proceed blindly, we used all the solvent molecules as input, without favoring those close to the ions in any way. To figure out whether to construct the graph using only oxygens or also hydrogens, we trained two CVs using the data collected in an unbiased simulation trajectory generated at 550 \unit{\kelvin}, one using all atoms to construct the graph and the other using only the heavy atoms. 

The GNN CV was optimized with the DeepTICA loss with a lag time of 0.2 \unit{\ps}\change{, which, as before, is the minimal value that will generate TICA eigenvalues larger than 0.85.}
According to the TICA theory, given the same lag time, CVs with a larger eigenvalue are associated with slower transition processes and will give a better description of the long-time dynamics of the system.~\cite{parrinello20211}
Therefore, we compared the eigenvalues of different MLCVs evaluated in their last validation epoch to assess their relative qualities.
We found that the eigenvalues of MLCVs trained without hydrogen atoms (0.98$\pm$0.01) are systemically larger than those of MLCVs trained with hydrogen atoms (0.91$\pm$0.01).
Such an observation hints that under our current model architecture, including hydrogen atoms in the MLCV will reduce its performance in learning the slow modes.
To further demonstrate this deduction, we evaluated the values of each MLCV as a function of the inter-ion distance $d_{\rm{NaCl}}$ (Figure~S2).
We observed that, although all MLCVs could clearly distinguish the associated and dissociated states, MLCVs trained with hydrogen atoms are suboptimal for recognizing the TS conformations.
Thus, we decided to use the heavy atom-based MLCV ($MLCV_{\rm{GNN}}^2$ in Figure~S2) to drive the following calculations.

To investigate the detailed behavior of the MLCV, we projected its values on the space spanned by $d_{\rm{NaCl}}$ and the oxygen coordination number of \ce{Na^{+}} ($CN_{\mathrm{NaO}}$), two important descriptors in the \ce{NaCl} dissociation process.~\cite{zhang2020}
From Figure~\ref{fig:2}A, it is clear that the change of the MLCV is dominated by $d_{\rm{NaCl}}$, while $CN_{\mathrm{NaO}}$ contributes to the CV value by a larger ratio in the dissociated state (the $d_{\rm{NaCl}} > 0.4\ \unit{\nm}$ region).
The node sensitivities of different species also support this observation: the CV is more sensitive to the ions than \change{any single} oxygen atom in the transition state region and tends to be equally sensitive to both species inside the dissociated state (Figure~\ref{fig:2}B).
Nevertheless, we can still tell that the oxygen atoms have played a role in the early dissociation process since their sensitivities are also considerably significant in the TS region.
We found that in this region, the average distance between the most sensitive oxygen atom and \ce{Na^{+}} is 0.25$\pm$0.07 \unit{\nm}, while the shortest distance between the oxygen atoms and \ce{Na^{+}} is about 0.22 \unit{\nm}.
\change{On the other hand, if we look at the cumulative sensitivity of the six oxygen atoms from the first coordination shell (red line in Figure~\ref{fig:2}B), we found that the overall importance of these atoms is still much higher in the TS region.}
Apparently, the oxygen atoms belonging to the first solvation shell of \ce{Na^{+}} are important to the CV, and it is thus reasonable to think that the CV is aware of the inner-shell water rearrangement during the dissociation.~\cite{jung20231}
\change{Interestingly, as shown in Figure~S3C, the cumulative sensitivity of all oxygen atoms in the box also exhibits a similar behavior along $d_{\rm{NaCl}}$.
Such a phenomenon further confirms that only ion-coordinated waters contribute to the CV, which is the desired behavior of a CV model for general ion-solvent systems.}

\change{We also applied the LASSO analysis to the CV model using the following four descriptors: $d_{\rm{NaCl}}$, $CN_{\mathrm{NaO}}$, hydrogen coordination number of \ce{Cl^{-}} ($CN_{\mathrm{ClH}}$) and number of the ``bridge'' waters connecting the two ions ($N_\mathrm{B}$).~\cite{ryan2014}
According to such analysis, the MLCV could be approximated by the following expression: $$MLCV\approx 0.86d_{\rm{NaCl}}+0.069N_\mathrm{B}+0.064CN_{\mathrm{NaO}}+0.007CN_{\mathrm{ClH}}.$$
As expected, the distance $d_{\rm{NaCl}}$ is found to be more important than the other descriptors.
Still, the summation of the contributions from $CN_{\mathrm{NaO}}$ and $N_\mathrm{B}$ is not negligible, confirming 
the important roles of water rearrangement in the ion association process.}

\if\ispreprint1
    \begin{figure*}[th!]\centering
    \includegraphics[width=\linewidth]{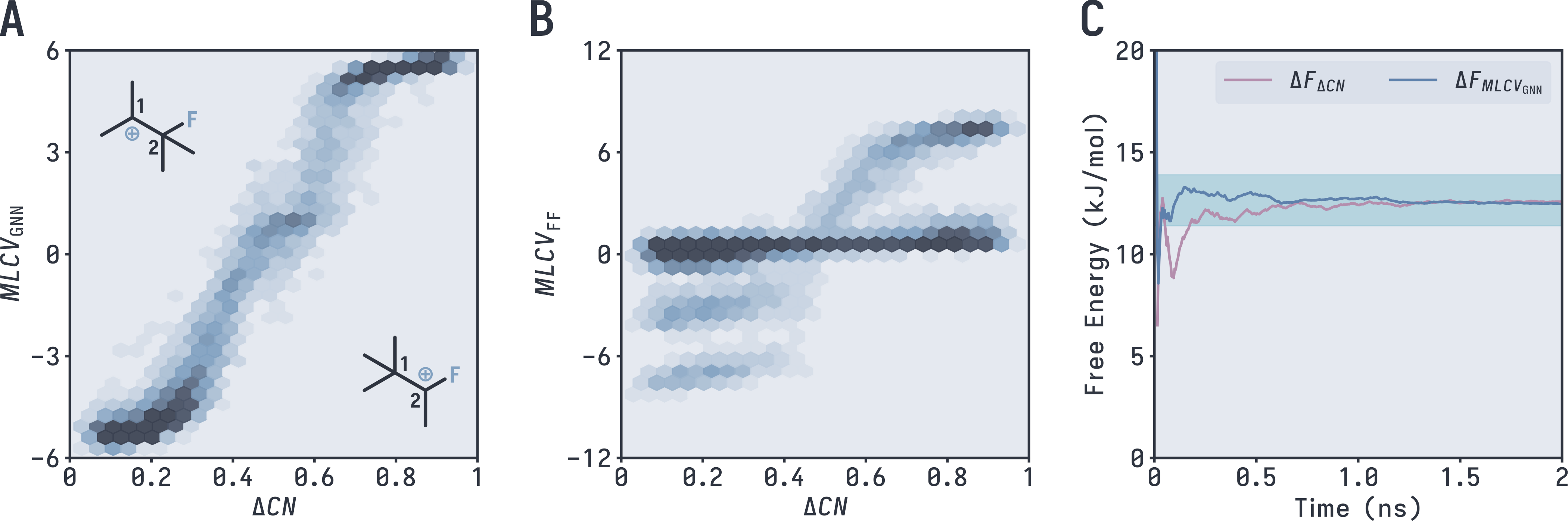}
    \caption{Results for the methyl migration process of FDMB cation. (A) Values of the GNN-based MLCV as a function of the $\Delta CN$ CV. The color of the cells stands for the density of conformations in the dataset. (B) Values of the feedforward network-based MLCV as a function of the $\Delta CN$ CV. The color of the cells stands for the density of conformations in the dataset. (C) The free energy difference between the two conformation states during the OPES simulation. Results from both the MLCV and classical CV-based simulations are shown. The reference free energy $\pm 0.5 k_{\mathrm{B}}T$ range is represented by the colored region.}
    \label{fig:3}
    \end{figure*}
\else
    \begin{figure}\centering
    \includegraphics[width=\linewidth]{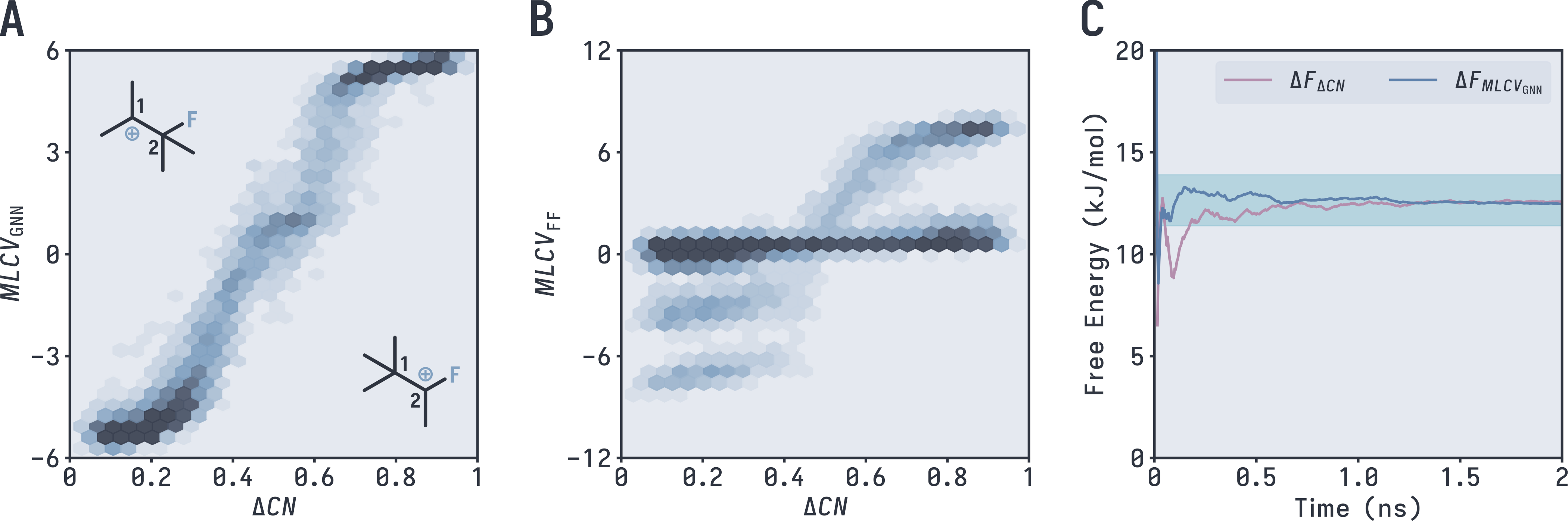}
    \caption{Results for the methyl migration process of FDMB cation. (A) Values of the GNN-based MLCV as a function of the $\Delta CN$ CV. The color of the cells stands for the density of conformations in the dataset. (B) Values of the feedforward network-based MLCV as a function of the $\Delta CN$ CV. The color of the cells stands for the density of conformations in the dataset. (C) The free energy difference between the two conformation states during the OPES simulation. Results from both the MLCV and classical CV-based simulations are shown. The reference free energy $\pm 0.5 k_{\mathrm{B}}T$ range is represented by the colored region.}
    \label{fig:3}
    \end{figure}
\fi

Finally, we assessed the performance of the MLCV in enhanced sampling.
We found that the free energy difference between the initial conformation state ($d_{\rm{NaCl}} < 0.35 \unit{\nm}$) and the final conformation state ($0.35 \unit{\nm} < d_{\rm{NaCl}} < 0.6 \unit{\nm}$) converged within $\pm 0.5 k_{\mathrm{B}}T$ of the reference value in a short simulation time (Figure~\ref{fig:2}C), and the sampling error estimated from independent OPES runs is also small (Table~\ref{tbl:1}).
Besides, the free energy profile calculated from the MLCV-based OPES simulations is also very similar to the one calculated from the reference OPES simulation based on $d_{\rm{NaCl}}$ and $CN_{\mathrm{NaO}}$ (Figure~S3).

It is worth noting that, compared with the number of oxygen atoms that actually contribute to the dissociation (the ones within 0.6 \unit{\nm} of \ce{Na^{+}}), the total number of oxygen atoms in the dataset is approximately six times more.
Thus, the training dataset of the MLCVs is considerably noisy and contains many irrelevant fast modes.
Even so, the network still successfully identified the central part of the system and accomplished reasonable descriptions of the dissociation process.
This gives confidence in applying the GNN-based MLCV with crudely constructed datasets.

\subsection{Methyl Migration Of FDMB Cation.}

As a third test for our method, we studied the methyl migration of the FDMB cation in a vacuum \changeR{(Figure~\ref{fig:0}C)} at 300 \unit{\kelvin}.
For the migration process, the four methyl groups in the molecule are equivalent (see inset of Figure~\ref{fig:3}A).
Thus, a good CV for this process should be invariant \textit{w.r.t.} the permutations of these methyl groups.
Fortunately, due to the permutation invariant nature of our GNN-based CV model, to train satisfactory MLCVs, we don't need to perform data augmentation by including permuted versions of the conformations in the dataset.
With this in mind, we built a dataset containing conformations collected from the reactant state, the product state, and the transition state.
Specifically, only heavy atoms in the cation were included in this dataset.
Using this dataset, we trained a GNN-based MLCV model by optimizing a three-state Transition Path Informed DeepTDA (TPI-DeepTDA) loss.~\cite{parrinello20233}
\change{Reactant and product conformations were sampled via unbiased MD starting from each state, while transition ones from biased simulations, either using some simple CVs or a DeepTDA CV trained on reactants/products (\change{see Section 1.2.3 of the SI for details}). The latter strategy ensures that the protocol can be used without requiring any advance information about transition pathways or appropriate CVs. } 

As a comparison, we also trained a feedforward network-based MLCV by optimizing the same loss function and using the distances between heavy atoms as the descriptors.
To compare the behaviors of the two different MLCVs, we plotted their values as functions of a simple permutation invariant CV: $\Delta CN = CN_{\mathrm{C_1C_{methyl}}} - CN_{\mathrm{C_2C_{methyl}}}$, which stands for a combination of the coordination numbers between the methyl carbon atoms and the two center carbon atoms (Figure~\ref{fig:3}A and Figure~\ref{fig:3}B).
The dataset used to evaluate the MLCVs is a biased simulation trajectory where the methyl groups are fully permuted.
From Figure~\ref{fig:3}A, it is clear that the GNN-based MLCV increases nearly monotonically with the increase of the $\Delta CN$ variable and is able to distinguish different conformation states.
In contrast, the feedforward network-based MLCV contains huge degeneracy (Figure~\ref{fig:3}B) and thus cannot be used in enhanced sampling.
To further prove that such degeneracy is caused by the lack of permutation invariance, we trained another feedforward network-based MLCV using conformations from this fully permuted trajectory and plotted its values again as a function of $\Delta CN$ (see Figure~S4).
We found that using the augmented dataset, the feedforward network-based MLCV exhibits the same behavior as the GNN-based MLCV.
Thus, there is no doubt that the intrinsic invariance of geometric GNN models could be very convenient for CV fitting tasks.

Then, we assessed the performance of the GNN-based MLCV in enhanced sampling.
We found that the free energy difference between the initial conformation state ($\Delta CN < 0.5$) and the final conformation state ($\Delta CN > 0.5$) converged within $\pm 0.5 k_{\mathrm{B}}T$ from the reference value in a short simulation time (Figure~\ref{fig:3}C), and the sampling error estimated from independent OPES runs is also small (Table~\ref{tbl:1}).
Again, these results exhibit the robustness of the GNN-based MLCV in biased simulations.

\section{Conclusions}

This work introduced a possible way of using geometric graph neural networks to build descriptor-free machine learning collective variables.
Relying on the invariance and great expression power of geometric GNNs, the proposed methodology can be applied to build robust CV models for various atomistic systems, without defining descriptors in advance.
Moreover, due to our methodology's generality, we can optimize the CV models according to diverse types of loss functions, e.g., classifiers and slow modes-based methods.

\changeR{The attribute of descriptor-free could also be given to other MLCVs that are based on distance (or contact) matrices, like the permutationally invariant networks for enhanced sampling (PINES)~\cite{ferguson2023} and DeepCV~\cite{ketkaew2022}, which also do not require selecting the descriptors manually.
However, we would argue that GNN-based CV models are more expressive and flexible, which means that one can adapt the GNN architectures to the required necessities.}

To assess the performance of our procedure, we fitted GNN-based MLCVs for three distinct testing systems, namely, alanine dipeptide in a vacuum, \ce{NaCl} in bulk water, and the FDMB cation in a vacuum.
These MLCVs were used to perform biased free energy simulations for each system.
We found that the MLCV-based simulations could achieve high accuracies in very short simulation time, which means that the MLCVs are optimal for free energy calculations.
Besides, we showed that we could easily obtain physical insights into the described processes from the CV models through node-level sensitivity analysis and the LASSO analysis.
Overall, our results exhibit the ability of geometric GNN-based MLCVs to capture key features of various atomistic processes.

Future directions include, from a technical standpoint, more efficient implementations of the graph construction and the study of the effect of different orders of equivariance/invariance on the system. 
In addition, we believe this methodology will be important for learning other important quantities useful for enhanced sampling simulations, from quantum-mechanical properties~\cite{bonati2023role} to the committor probability,~\cite{parrinello2024, trizio2024onceprobabilitybasedenhancedsampling} bringing us one step closer to universal CVs. 

\section*{Acknowledgements}
This work received support from the National Natural Science Foundation of China (22220102001 and 92370130).
L.B. and M.P. acknowledge funding from the Federal Ministry of Education and Research, Germany, under the TransHyDE research network AmmoRef (support code: 03HY203A).
J. Z. would like to thank Dhiman Ray and Peilin Kang for many beneficial discussions.

\section*{Data availability}

The code to train the CVs and perform the simulations to reproduce the results presented here are available on GitHub: \url{https://github.com/jintuzhang/gnncv}.

The modified \verb|mlcolvar| package \change{supporting GNNs} is also available on GitHub at: \url{https://github.com/jintuzhang/mlcolvar} \change{and we plan to port the new features in the official }\verb|mlcolvar| ~\change{library}~\cite{parrinello20232}.

\section*{Bibliography}
\bibliography{main}
\end{document}


\section{Computational Details.}

\subsection{Network Architecture And Hyperparameters.}

We adopted the GVP-GNN architecture proposed by \citeauthor{dror20211}.
On top of that, instead of the original Gaussian basis set, we used a set of Bessel functions multiplied by a smooth polynomial cutoff to expand the interatomic distances,~\cite{gasteiger2022} to improve the smoothness of the CV models.
For training the networks, we used the Adam optimizer with an exponential learning rate scheduler.
We listed detailed hyperparameters of the network and the optimizer for each system in Table~\ref{tbl:s1} and Table~\ref{tbl:s2}, correspondingly.

Besides, for the feedforward network-based CV model used to describe the methyl migration of FDMB cation, we used a neural network with one output and a 21-48-12-1 architecture.
Other hyperparameters for training the network were kept the same as the ones used for training the GNN-based CV model, as listed in Table~\ref{tbl:s2}.
Especially, for the TDA loss functions used in studying the FDMB cation system, we selected the hyperparameters $\alpha$, $\beta$, $\sigma$ and $\mu$ as 1.0, 100.0, \{-7, 0, 7\} and \{0.5, 1, 0.5\}.
The later parameter were selected \textit{w.r.t.} the principle introduced by~\citeauthor{parrinello20223}.

\begin{table}
\caption{Network hyperparameters for each system.}
\label{tbl:s1}

\begin{tabular}{cccccccc}
    \hline
    System &
        \texttt{cutoff} &
        \texttt{n\_layers} &
        \texttt{n\_out} &
        \texttt{n\_bases} &
        \texttt{n\_polynomials} &
        \texttt{n\_messages} &
        \texttt{n\_ff} \\
    \hline Ala2      & 1.0 & 2 & 6 & 10 & 6 & 2 & 2 \\
    \hline \ce{NaCl} & 0.6 & 2 & 8 &  8 & 6 & 2 & 2 \\
    \hline FDMB      & 0.8 & 1 & 1 &  8 & 6 & 2 & 2 \\
    \hline
\end{tabular}

\texttt{cutoff}: Cutoff radius of the basis function and the neighbor list, in a unit of \unit{\nm}. \\
\texttt{n\_layers}: Number of GVP convolution layers in the network. \\
\texttt{n\_out}: Output feature dimensionality of the network. \\
\texttt{n\_bases}: Number of Bessel basis functions. \\
\texttt{n\_polynomials}: Order of the smooth cutoff polynomial. \\
\texttt{n\_messages}: Number of message functions in each convolutional layer. \\
\texttt{n\_ff}: Number of feedforward functions in each convolutional layer.

\end{table}

\begin{table}
\caption*{Table~\ref{tbl:s1} Continued.}

\begin{tabular}{cccccc}
    \hline
    System &
        \texttt{n\_scalars\_node} &
        \texttt{n\_vectors\_node} &
        \texttt{n\_scalars\_edge} &
        \texttt{r\_drop} &
        \texttt{activation} \\
    \hline Ala2      & 6 & 2 & 6 & 0.0 & \texttt{Tanh} \\
    \hline \ce{NaCl} & 6 & 4 & 6 & 0.0 & \texttt{Tanh} \\
    \hline FDMB      & 8 & 8 & 8 & 0.2 & \texttt{SiLU} \\
    \hline
\end{tabular}

\texttt{n\_scalars\_node}: Size of the scalar channels of the node embeddings.\\
\texttt{n\_vectors\_node}: Size of the vector channels of the node embeddings.\\
\texttt{n\_scalars\_edge}: Size of the scalar channels of the edge embeddings.\\
\texttt{r\_drop}: Probability of the \texttt{dropout} layers (the drop rate). \\
\texttt{activation}: Type of the activation functions. \\

\end{table}

\begin{table}
\caption{Optimizer hyperparameters for each system.}
\label{tbl:s2}

\begin{tabular}{cccccc}
    \hline
    System &
        Batch Size &
        Train/Valid Ratio &
        Initial LR &
        Weight Decay &
        LR Scheduler $\gamma$ \\
    \hline Ala2      & 5000 & 4:1 & 3E-3 & 2E-4 & 0.9995 \\
    \hline \ce{NaCl} & 1000 & 4:1 & 4E-3 & 1E-4 & 0.9997 \\
    \hline FDMB      & 3000 & 4:1 & 4E-3 & 1E-4 & 0.9997 \\
    \hline
\end{tabular}
\end{table}

\subsection{Molecular Dynamics Simulations.}

\subsubsection{Alanine Dipeptide.}

We utilized the AMBER99-SB~\cite{ross2013} force field to describe the alanine dipeptide molecule in a vacuum.
The lengths of all chemical bonds involving hydrogen atoms were constrained at their equilibrium values using the RATTLE~\cite{andersen1983} algorithm.
The electrostatic and Lennard-Jones (LJ) interactions were described in a non-cutoff manner.
All MD propagations were carried out under the \textit{NVT} ensemble using the OpenMM~\cite{openmm} package with the Geodesic BAOAB Langevin integrator~\cite{leimkuhler2016} provided by OpenMMTools.~\cite{openmmtools}
The integration time step and the friction coefficient were set to 2 \unit{\fs} and 1 \unit{\ps^{-1}}, correspondingly.
The number of geodesic drift steps was set to 2.

To build the training set for the MLCV, we carried out a \char`\~5.5 \unit{\ns} MD run starting from the $C_\mathrm{5}$ conformation, at a temperature of 600 \unit{\kelvin}.
As a result, we obtained an unlabeled dataset containing 27,281 conformations.
After obtaining the MLCV, we performed three independent 20 \unit{\ns} On-the-fly Probability Enhanced Sampling (OPES)~\cite{parrinello20202} simulations at 300 \unit{\kelvin} by biasing the MLCV.
In these MLCV-based OPES simulations, we used a \texttt{BARRIER} parameter of 35 \unit{\kJ/\mole}, and the kernel functions were deployed every 500 steps.
To compare with, a 30 \unit{\ns} reference OPES simulation was also carried out, where the two torsional angles $\phi$ and $\psi$ were used as the CVs.
In this torsional angles-based OPES simulation, we used a \texttt{BARRIER} parameter of 45 \unit{\kJ/\mole}, and the kernel functions were deployed every 500 steps as well.
All the free energy calculations were performed with the community-developed plugin for molecular dynamics (PLUMED) \cite{plumed}, version 2.9.0.

\subsubsection{\ce{NaCl} Dissociation In Water.}

We utilized the CHARMM22~\cite{mackerell2004} force field with the ion parameters presented by~\citeauthor{beglov1994} to describe the \ce{NaCl} solution system.
In our setup, one \ce{NaCl} ion pair was solvated to a density of 1 \unit{g/cm^3} with 216 TIP3P~\cite{jorgensen1983} waters in a $1.86 \times 1.86 \times 1.86$ \unit{\nm^3} cubic box.
The lengths of all \ce{H-O} bonds were constrained at their equilibrium values using the RATTLE~\cite{andersen1983} algorithm.
The electrostatic interactions were described using the Particle Mesh Ewald (PME) \cite{essmann1995} method with a real-space cutoff at 0.9 nm.
The Lennard-Jones (LJ) interactions were calculated with a distance cutoff of 0.9 nm,
All MD propagations were carried out under the \textit{NVT} ensemble using the OpenMM~\cite{openmm} package with the Geodesic BAOAB Langevin integrator~\cite{leimkuhler2016} provided by OpenMMTools.~\cite{openmmtools}
The integration time step and the friction coefficient were set to 2 \unit{\fs} and 1 \unit{\ps^{-1}}, correspondingly.
The number of geodesic drift steps was set to 4.

To build the training set for the MLCV, we carried out a 2 \unit{\ns} MD run starting from the bounded conformation, at a temperature of 550 \unit{\kelvin}.
During the unbiased MD run, we restrained the maximum value of the distance between the two ions ($d_{\mathrm{NaCl}}$) by placing an \texttt{UPPER\_WALLS} at the distance of 0.8 \unit{\nm}.
The force constant of the wall was selected as 2000 \unit{\kJ/\mole/\nm^2}.
As a result, we obtained an unlabeled dataset containing 20,000 conformations.
In particular, we selected the cutoff radius of constructing the graphs as 0.6 \unit{\nm}, which is the approximate size of the second solvent shell of the \ce{Na^+} ion.

After obtaining the MLCVs, we then performed three independent 15 \unit{\ns} OPES simulations at 300 \unit{\kelvin} by biasing one of the MLCVs that only contain the heavy atoms ($MLCV_{GNN}^2$ in Figure~\ref{fig:s2}).
Meanwhile, a 100 \unit{\ns} reference OPES simulation was carried out, by biasing $d_{\mathrm{NaCl}}$ and the oxygen coordination number of the sodium ion ($CN_{\mathrm{NaO}}$).
The definition of $CN_{\mathrm{NaO}}$ follows the one suggested by \citeauthor{zhang2020}.
In all the above OPES simulations, we used a \texttt{BARRIER} parameter of 15 \unit{\kJ/\mole}, and the kernel functions were deployed every 500 steps as well.
Besides, in these simulations, we again constrained the maximum value of the distance between the two ions ($d_{\mathrm{NaCl}}$) by placing an \texttt{UPPER\_WALLS} at the distance of 0.6 \unit{\nm}.
The force constant of the wall was selected as 2000 \unit{\kJ/\mole/\nm^2}.
All the biased simulations were performed with the community-developed plugin for molecular dynamics (PLUMED)~\cite{plumed}, version 2.9.0.

In order to count numbers of the ``bridge'' waters within the two ions for performing the LASSO analysis, we applied the $N_\mathrm{B}$ descriptor introduced by~\citeauthor{ryan2014}.
Besides, for the hydrogen coordination number of the chloride ion ($CN_{\mathrm{ClH}}$), we used the same coordination number definition as that of $CN_{\mathrm{NaO}}$~\cite{zhang2020}, and selected the following values for the \verb|R_0|, \verb|D_0|, \verb|MM| and \verb|NN| parameters: 0.1, 0.205, 4, 12.~\cite{ryan2014}

\subsubsection{Methyl Migration Of FDMB Cation.}

We utilized the GFN1-xTB~\cite{grimme2017} semiempirical tight binding method to simulate the methyl migration of an FDMB cation in a vacuum.
All MD propagations were carried out under the \textit{NVT} ensemble using the SOMD~\cite{somd} package with the BAOAB Langevin integrator~\cite{leimkuhler2013}.
The integration time step and the friction coefficient were set to 0.5 \unit{\fs} and 20 \unit{\ps^{-1}}, correspondingly.

To build the training set for the MLCVs, a 2.5 \unit{\ns} reference OPES simulation was carried out at 300 \unit{\kelvin}.
In the reference OPES simulation, a combination of the coordination numbers between the methyl carbon atoms and the two center carbon atoms: $\Delta CN = CN_{\mathrm{C_1C_{methyl}}} - CN_{\mathrm{C_2C_{methyl}}}$ was used as the CV.
The \texttt{R\_0}, \texttt{NN}, and \texttt{MM} parameters for defining the coordination numbers were selected as 0.16 \unit{\nm}, 6 and 12, correspondingly.
For the biasing potential, we used a \texttt{BARRIER} parameter of 30 \unit{\kJ/\mole}, and the kernel functions were deployed every 500 steps.
To prevent the system from entering nonphysical conformations, we restrained the minimum and maximum values of the $\Delta CN$ CV by placing a \texttt{LOWER\_WALLS} and an \texttt{UPPER\_WALLS} at the positions of 0.1 and 0.9, correspondingly.
The force constant of the walls was selected as 2000 \unit{\kJ/\mole/{cv\_unit}^2}.

From the reference biased simulation trajectory, we collected 1327 transition state (TS) conformations, which contain a $\Delta CN$ value between (0.4, 0.6).
Then we performed another two 50 \unit{\ps} unbiased simulations starting from the reactant and product structures, at a temperature of 300 \unit{\kelvin}.
Using the last 40 \unit{\ps} of both trajectories and the TS structures, we constructed a labeled dataset that contains 2927 conformations and three different state labels.
This dataset was used to train the GNN-based and the first feedforward network-based MLCV.

Specifically, we found that in the 2.5 \unit{\ns} reference OPES simulation, all the methyl groups had been fully permuted spontaneously.
Thus, we used this trajectory to evaluate the CV values shown in Figure~3.
Besides, we also collected the reactant structures (structures with a $\Delta CN$ value smaller than 0.15) and product structures (structures with a $\Delta CN$ value larger than 0.85) from this trajectory, to train the second feedforward network-based MLCV, which's values were shown in Figure~\ref{fig:s4}.

After obtaining the MLCVs, we performed three independent 2 \unit{\ns} OPES simulations driven by the GNN-based MLCV, during which the parameters used in the reference OPES simulation were applied again.

Here, we state that, the method of obtaining the TS structures is not limited to performing an auxiliary biased simulation using physical-informed CVs.
To this end, we trained another 2-state GNN-based DeepTDA CV using only the metastable state conformations.
Then, based on such a CV, we performed a 2 \unit{\ns} OPES simulations, and the transition of the $\Delta CN$ variable during this simulation is shown in Figure~\ref{fig:s6}.
As can be seen, during this simulation we could still obtain enough transitions between the reactant and product states within the same simulation time, as well TS region structures.
Thus, for realistic systems which are not able to generate transitions using simulating annealing, we may adopt the iterative CV fitting procedure reported by~\citeauthor{parrinello20211} to obtain robust CVs.

\section{Supplementary Figures}

\begin{figure}
    \includegraphics[width=\linewidth]{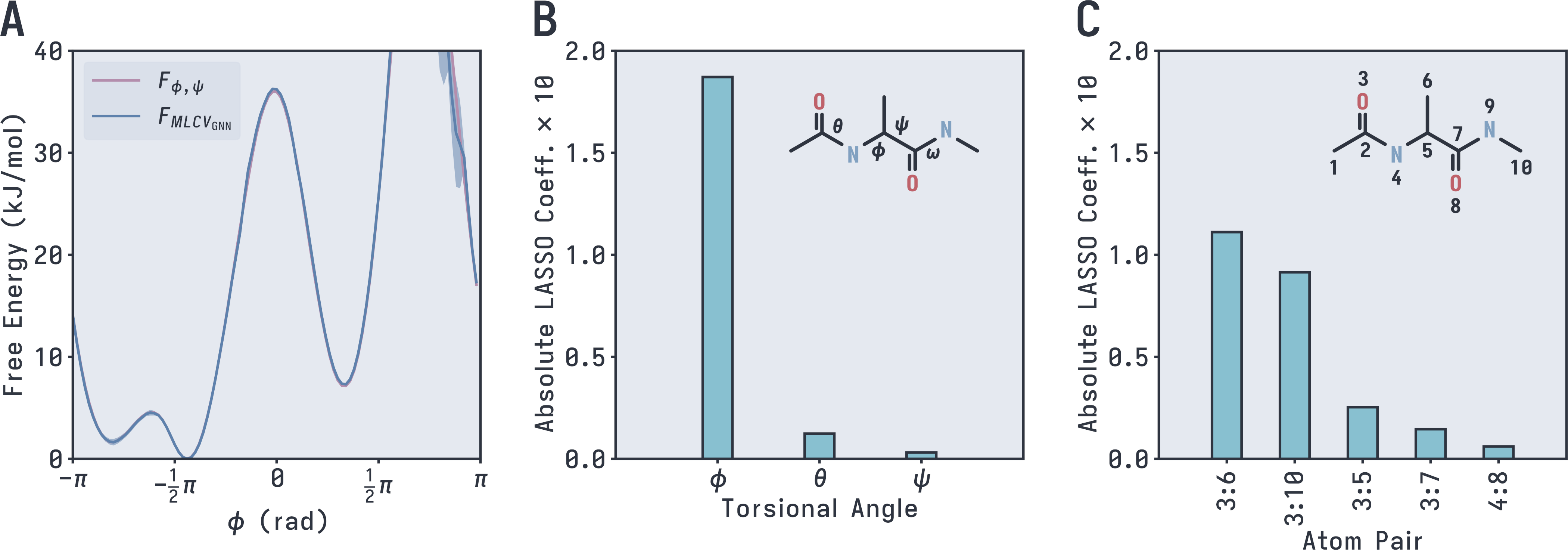}
    \caption{Extended results for the conformation transition of alanine dipeptide. (A) Free energy profile of the system projected on the $\phi$ torsional angle. Results from both the MLCV and the torsional angles-based OPES simulations are shown. (B) Absolute LASSO coefficients of the torsional angles. Before the analysis, the range of the MLCV has been scaled to [-1, 1]. The analysis suggests that the MLCV could be approximated as $MLCV \approx 0.187\phi + 0.012\theta - 0.003\psi$ (C) Absolute LASSO coefficients of the inter-atom distances. Before the analysis, the range of the MLCV has been scaled to [-1, 1].The analysis suggests that the MLCV could be approximated as $MLCV \approx - 0.111d_{3:6} - 0.091d_{3:10} + 0.025d_{3:5} - 0.015d_{3:7} + 0.006d_{4:8}$.}
    \label{fig:s1}
\end{figure}

\begin{figure}
    \includegraphics[width=\linewidth]{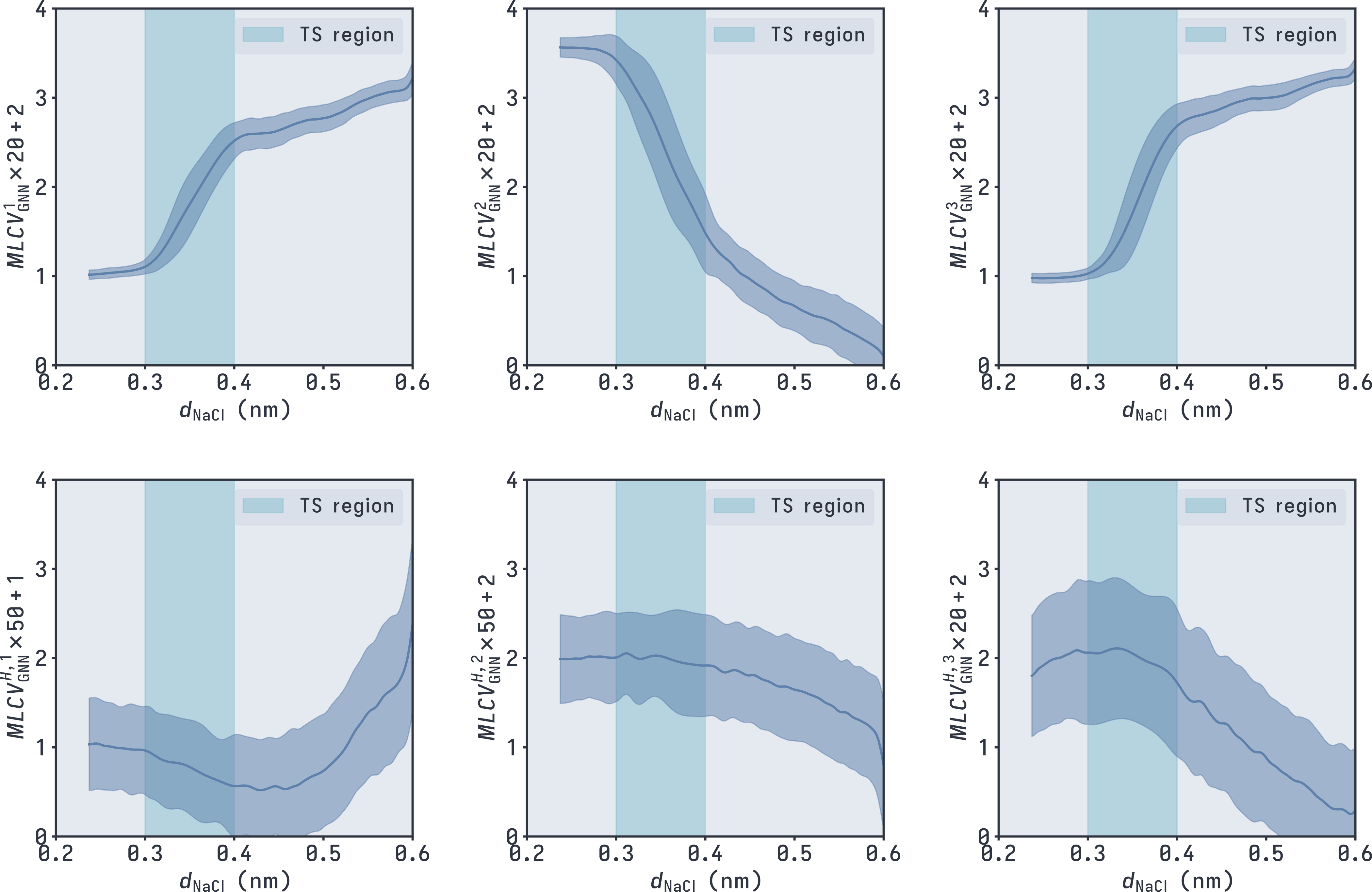}
    \caption{The values of each MLCV for the \ce{NaCl} dissociation process as a function of the inter-ion distance $d_{\rm{NaCl}}$. The MLCVs shown in the first row ($MLCV_{\mathrm{GNN}}^1$ to $MLCV_{\mathrm{GNN}}^3$) are trained without hydrogens, and the MLCVs shown in the second row ($MLCV_{\mathrm{GNN}}^{H,1}$ to $MLCV_{\mathrm{GNN}}^{H,3}$) are trained with hydrogens}
    \label{fig:s2}
\end{figure}

\begin{figure}
    \includegraphics[width=\linewidth]{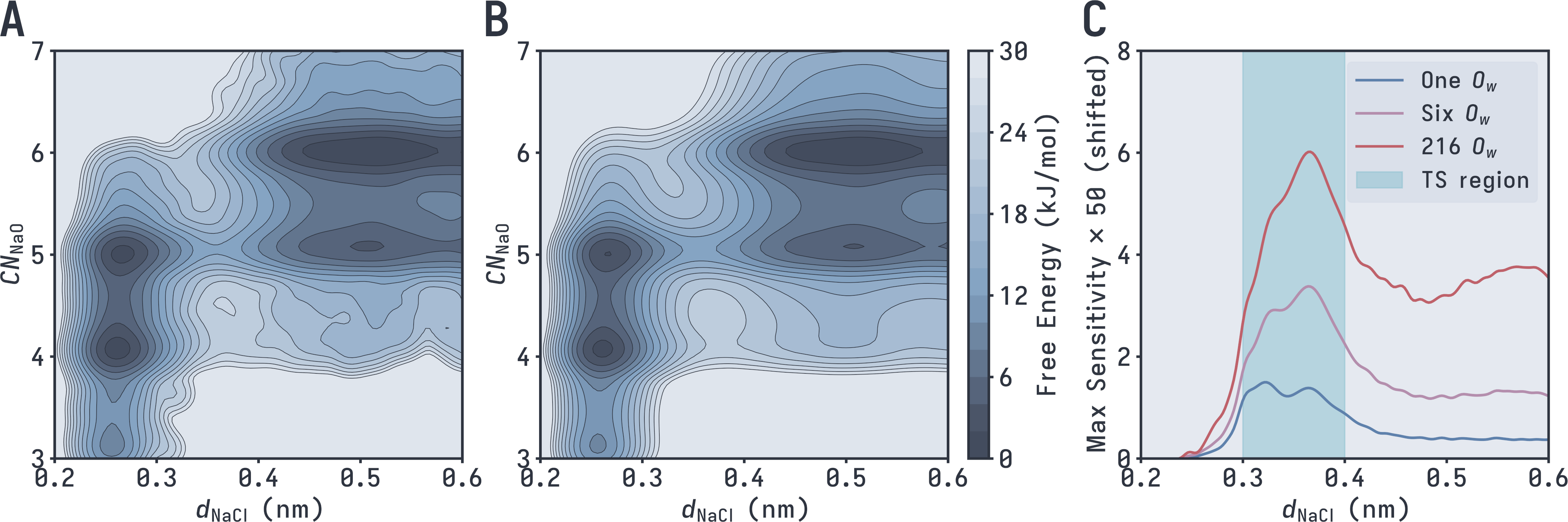}
    \caption{Free energy profile of the \ce{Nacl} dissociation process projected on $d_{\mathrm{NaCl}}$ and $CN_{\mathrm{NaO}}$. (A) Result from the MLCV-based OPES simulations. (B) Result from the physical CVs-based OPES simulation. (C) Changing of max sensitivity of the different number of oxygen atoms \textit{w.r.t.} $d_{\rm{NaCl}}$. The minimal values of all curves are shifted to zero.}
    \label{fig:s3}
\end{figure}

\begin{figure}
    \includegraphics[width=0.5\linewidth]{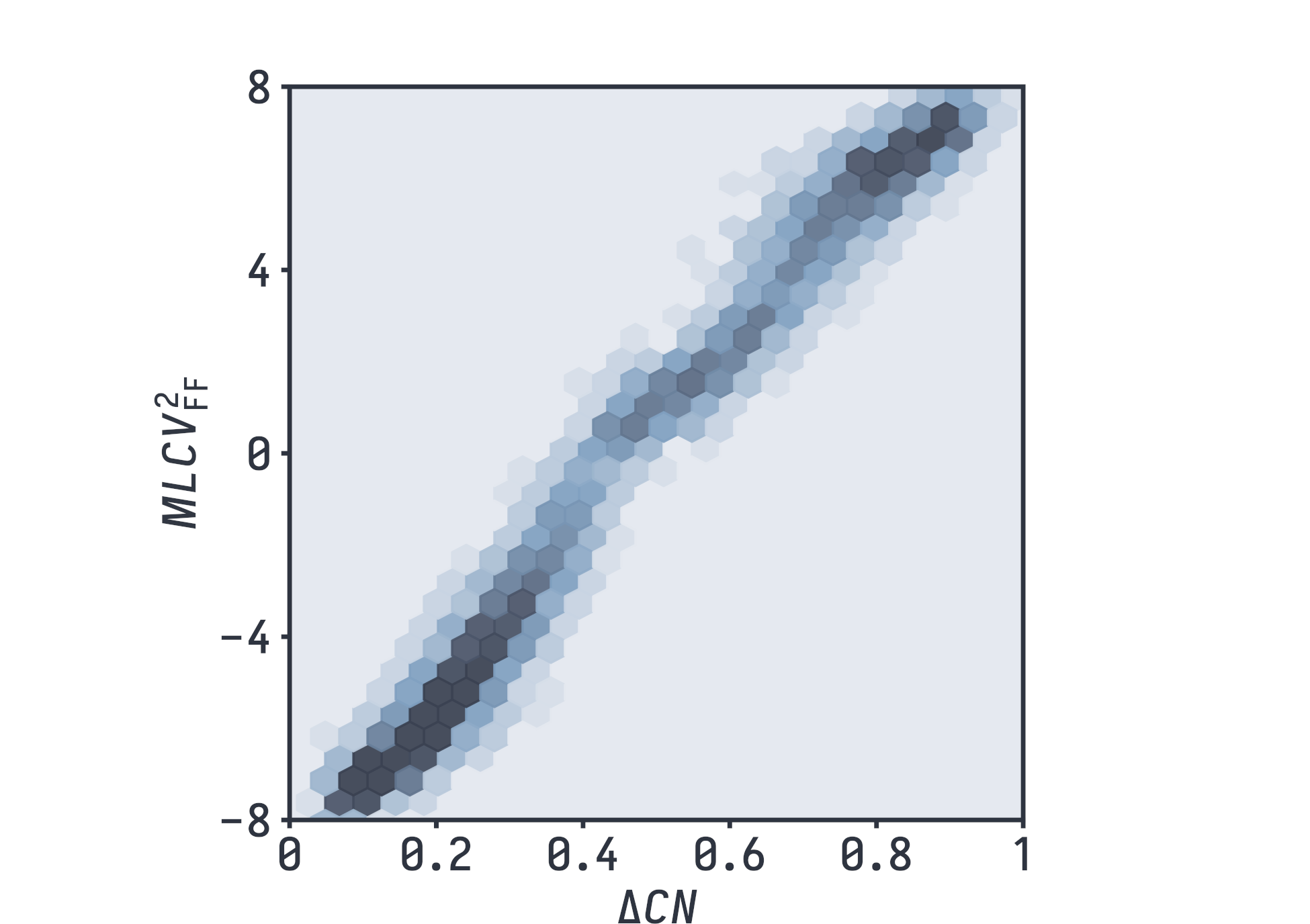}
    \caption{Values of the feedforward network-based MLCV for the methyl migration process of FDMB cation as a function of $\Delta CN$. This CV model was trained using the dataset where all the methyl groups were fully permuted.}
    \label{fig:s4}
\end{figure}

\begin{figure}
    \includegraphics[width=\linewidth]{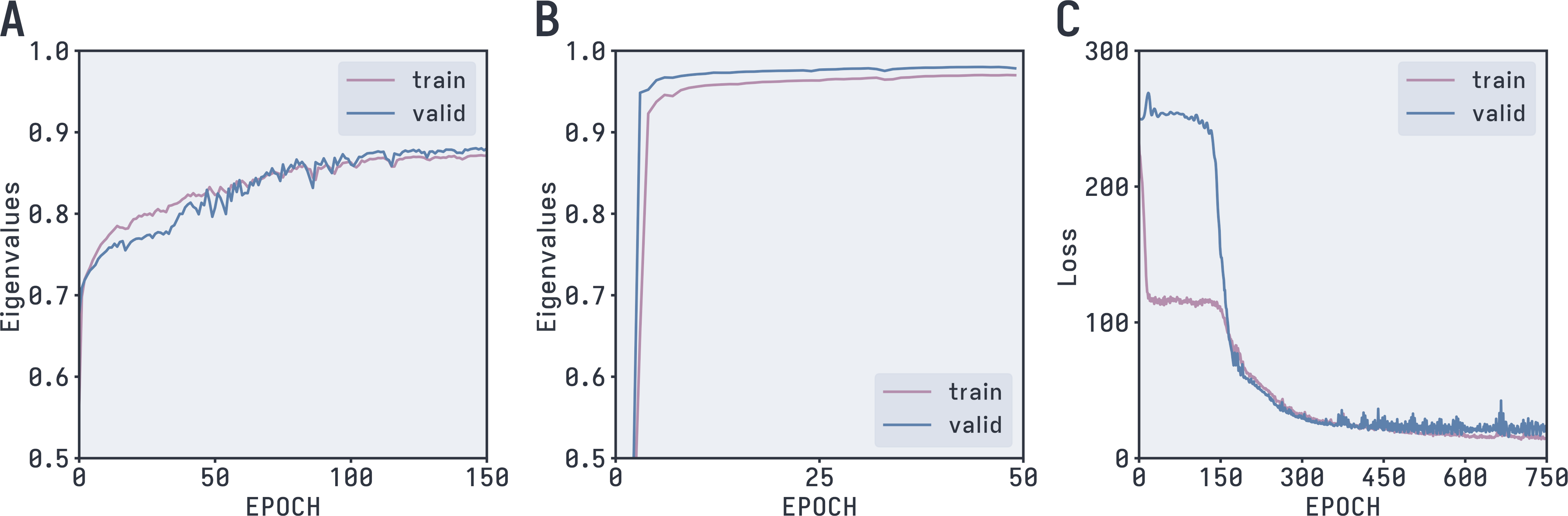}
    \caption{Typical loss curve for the three studied systems. (A) Alanine dipeptide. (B) \ce{NaCl} Dissociation In Water. (C) Methyl Migration Of FDMB Cation.}
    \label{fig:s5}
\end{figure}

\begin{figure}
    \includegraphics[width=0.5\linewidth]{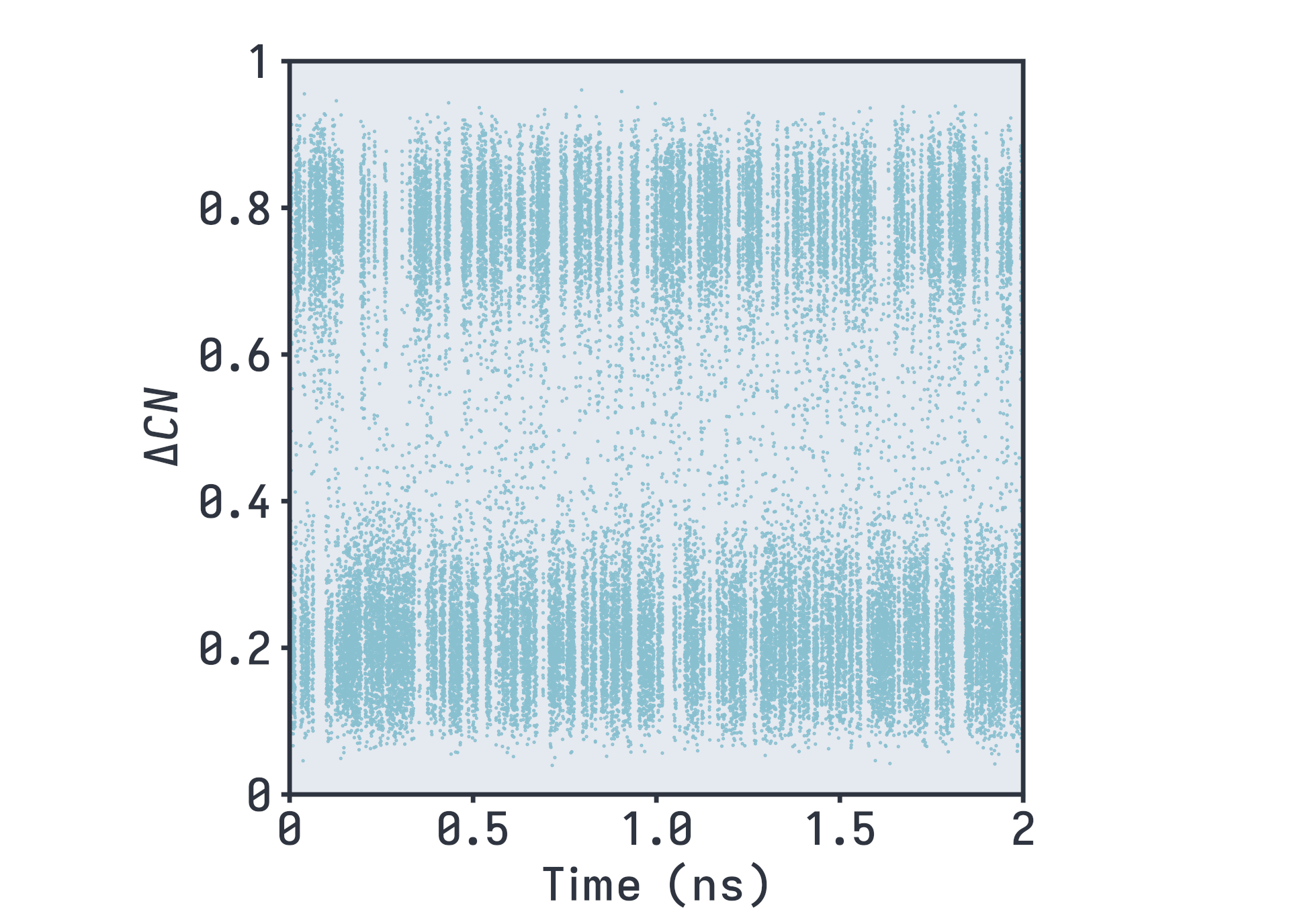}
    \caption{Transitions of the $\Delta CN$ variable during this OPES simulation of the FDMB cation system driven by the 2-state GNN-based DeepTDA CV.}
    \label{fig:s6}
\end{figure}

\clearpage
\bibliography{main}